\documentclass[reprint,aps,prl,superscriptaddress,amsmath,amssymb,floatfix,footinbib,longbibliography]{revtex4-1}
\usepackage{graphicx}
\usepackage{xcolor}
\usepackage{bm}
\usepackage{ulem}
\usepackage[colorlinks, linkcolor= blue, citecolor = blue, urlcolor=blue]{hyperref}
\usepackage{comment}
\usepackage{physics}
\usepackage{float}
\usepackage{pifont}
\usepackage{wrapfig}
\usepackage[stable]{footmisc}
\usepackage{array}
\usepackage{gensymb}
\usepackage{multirow}
\usepackage{hhline}
\newcommand{\cmark}{\ding{51}}%
\newcommand{\xmark}{\ding{55}}%
\usepackage{chemformula}

\def\bea{\begin{eqnarray}}
\def\eea{\end{eqnarray}}
\def\be{\begin{equation}}
\def\ee{\end{equation}}

\def\bal{\begin{aligned}}
\def\eal{\end{aligned}}

\begin{document}
\title{Light-induced Nonlinear Resonant Spin Magnetization}

\author{Sayan Sarkar}
\author{Sunit Das}
\author{Debottam Mandal}
\author{Amit Agarwal}
\email{amitag@iitk.ac.in}
\affiliation{Department of Physics, Indian Institute of Technology, Kanpur-208016, India}

\begin{abstract}
The optical generation of nonequilibrium spin magnetization plays a crucial role in advancing spintronics, providing ultrafast control of magnetization dynamics without the need for magnetic fields. Here, we demonstrate the feasibility of light-induced nonlinear spin magnetization (LNSM), which becomes a dominant effect in centrosymmetric materials. We reveal the quantum geometric origins of various LNSM contributions in both metallic and insulating systems. Through detailed symmetry analysis, we predict significant LNSM in the antiferromagnetic material CuMnAs. Notably, under circularly polarized light, the spin magnetization exhibits helicity-dependent behavior, reversing with opposite light helicity. These findings open up new possibilities for generating LNSM-driven nonlinear spin-orbit torques and developing innovative opto-spintronic devices.

\end{abstract}

\maketitle

\textcolor{blue}{\it Introduction:--} The optical control of charge and spin in materials is vital for both fundamental physics and technological advancements~\cite{Song_nrp23,Schiffrin_nature2013,Reserbat_ACS2021,Bader_ARC2010,Liu_nature2020}. Optically generated nonlinear charge currents are widely used to probe topological and magnetic properties in materials, with applications in photovoltaic technologies~\cite{Ma_NM21}. The optical manipulation of spin degrees of freedom has shown great promise for ultrafast spintronic devices with minimal dissipation~\cite{HIROHATA20,Jungwirth_12,Bibes_RMP24}.

Controlling spin without magnetic fields, particularly through optical and electrical means, represents the frontier of spintronics research. This holds immense potential for high-density data storage, memory technologies, and high-speed computing~\cite{sharma_rmp04,HIROHATA20,Puebla_cm20}. Substantial research efforts have focused on controlling and manipulating the spin magnetization and spin-orbit torques in magnetic materials~\cite{manchon_rmp19,Sierra_21,Ting_am23,Manuel_rmp24}. The spin magnetization ($\delta {\bm S}$) induced by an electric field ($\bm E$) in the linear regime is described by $\delta S_a^{(1)} = \alpha_{a;b} E_b$, where $a,b$ are Cartesian coordinates. The linear spin response has been successfully used to control magnetic order in noncentrosymmetric materials~\cite{EDELSTEIN_90,Zelenzy_prb17,Zelenzy_prb17,Sanchez_nc13,Mellnik_Nat14,Nui_pnas20,Annika_pubmed24,Song_NC24}. However, in centrosymmetric systems, symmetry constraints force the linear spin magnetization to vanish~\cite{Zelezny_prl14,Zelenzy_prb17,Itou_prr21,Ilya_prb23,Yao_prbL24,Yang_arxiv24}, limiting its utility. 

\begin{figure}[t!]
    \centering
    \includegraphics[width=1.0\linewidth]{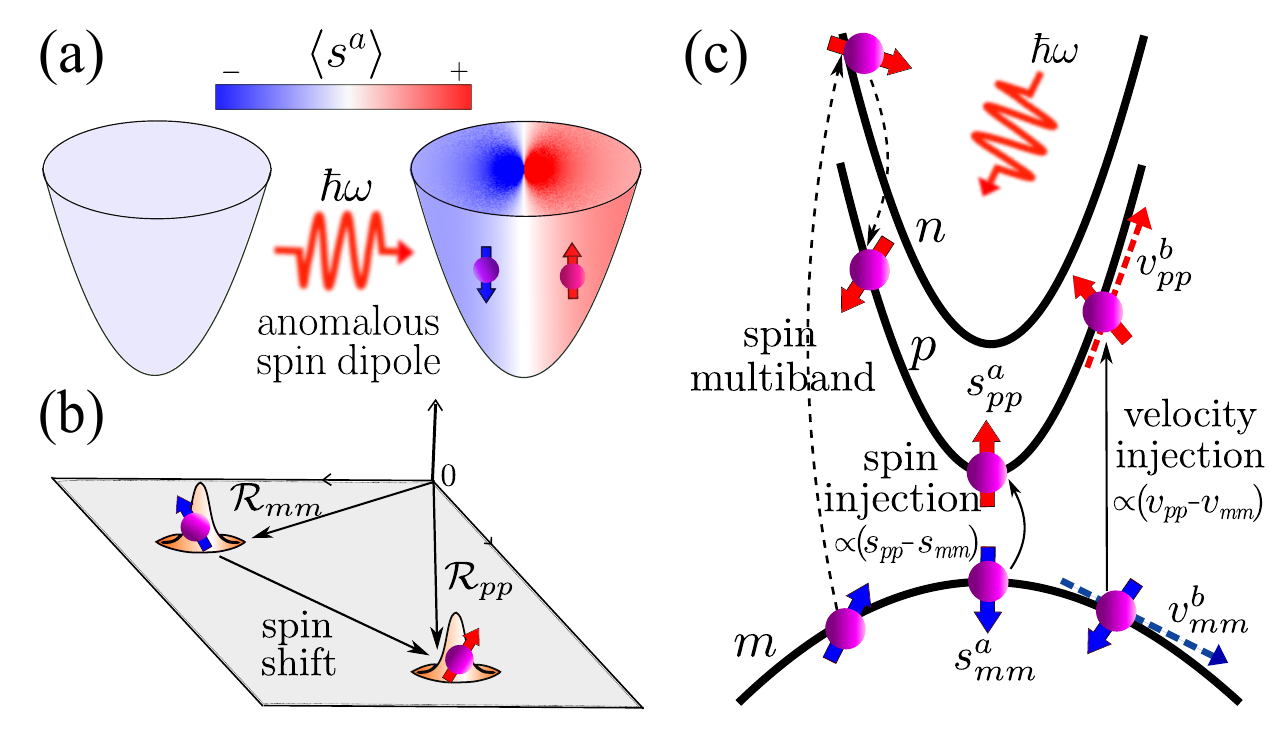}
    \caption{{\bf Band geometric origin of LNSM}. (a) The anomalous spin magnetization component arises due to the electric field-induced spin dipole moment. (b) The change in the spin orientation during the real-space shift of wave packets gives rise to the spin-shift magnetization. (c) The velocity injection and spin injection contributions to spin magnetization arise from the change in the velocities of electrons and the change of spin expectation during transitions between a pair of bands. Additionally, multiple interband transitions of optically excited electrons generate the multiband spin magnetization.
    \label{Fig_1}}
\end{figure}

In this Letter, we present novel mechanisms for light-induced nonlinear spin magnetization (LNSM), which overcomes these limitations and offers a robust way of controlling spin magnetization using light polarization. This second-order effect, described by $\delta S_a^{(2)} = \alpha_{a;bc} E_b E_c$, is not bound by the symmetry constraints that affect linear spin magnetization. As a result, it can occur in both centrosymmetric and noncentrosymmetric materials~\cite{Xiao_2022,Xiao_2023,Fregoso_2022,Zhou_nature22,fregoso_photo24,Li_prb21,HWLee_npj24}. We derive the key contributions to LNSM arising from the band geometry and explain their physical significance [see Fig.~\ref{Fig_1}]. Our comprehensive symmetry analysis identifies a wide range of materials where nonlinear spin magnetization can be significant. 

We demonstrate this novel phenomenon in antiferromagnetic \ch{CuMnAs}, where light polarization plays a pivotal role in driving spin magnetization. Under terahertz light illumination, we predict a large nonlinear spin magnetization in CuMnAs. Furthermore, when circularly polarized light is used, the LNSM in CuMnAs reverses its direction depending on the light’s helicity. This discovery provides a new pathway for generating nonlinear spin-orbit torques in centrosymmetric magnetic materials, enabling all-optical control of magnetization dynamics~\cite{Fang_24_perpective,Wu_23ultra,Hamamera_cp22,Wang_NConv20}.
LNSM has the potential to enable magnetization switching in ferromagnets and antiferromagnets, as well as in altermagnets~\cite{Gambardella2011, Jungwirth_prl14,Jungwirth_prb17, manchon_rmp19,Yang__24quantum,Wang_prb22}. Additionally, it could facilitate the development of coherent terahertz oscillators in antiferromagnets~\cite{Cheng_prl15, Khymyn_nature17,Song_NC24}. These findings open up exciting possibilities for designing energy-efficient, ultrafast opto-spintronic devices.

\textcolor{blue}{\it Light-induced spin magnetization:--} 
We calculate the dynamical nonlinear spin magnetization, utilizing the quantum kinetic theory  framework~\cite{Sekine_prb17}. The dynamics of Bloch electrons under an external field are described by the density matrix $\rho({\bm k},t)$, which evolves according to the quantum Liouville equation: $i \hbar \, \partial_t \rho({\bm k},t) = [{\cal H},\rho({\bm k},t)]$, with ${\cal H}={\cal H}_0+ {\cal H}_E$. Here, ${\cal H}={\cal H}_0+ {\cal H}_E$ comprises  of the unperturbed Bloch Hamiltonian ${\cal H}_0$ and the light-matter interaction Hamiltonian ${\cal H}_E$. The Bloch Hamiltonian ${\cal H}_0$ defines the Bloch states $|u_{m \bm k}\rangle$ with corresponding energies $\varepsilon_{m \bm k}$, satisfying ${\cal H}_0 |u_{m \bm k}\rangle =  \varepsilon_{m \bm k} |u_{m \bm k}\rangle$. The light-matter interaction is characterized by an external electric field $\bm{E}(t)$, and it is expressed as ${\cal H}_E = e \hat{\bm{r}}\cdot\bm{E}(t)$~\footnote{In principle, the magnetic field ${\bm B}(t)$ of the electromagnetic field should also be considered in the light-matter coupling term. However, the magnetic coupling is very weak compared to the electric field of the light as $B \sim E/v_0$, with $v_0$ being the velocity of light in the material in discussion. Consequently, the spin-Zeeman coupling term in the light-matter interaction Hamiltonian is very weak compared to the electric field-induced term.}. We assume a homogeneous monochromatic electric field $\bm{E}(t) = |{\bm E}| e^{-i\omega t}+\mathrm{c.c.}$, with $\bm{E}=\{E_x, E_y, E_z\}$ representing the field strength and $\omega$ the frequency of the incident light.

We solve the quantum Liouville equation perturbatively by expanding the density matrix $\rho$ in powers of the electric field amplitude $|\bm{E}|$, leading to $\rho = \rho^{(0)} + \rho^{(1)} + \rho^{(2)} + \cdots$, where $\rho^{(N)} \propto |\bm{E}|^N$. The zeroth-order term $\rho^{(0)}$ corresponds to a diagonal matrix with elements given by the equilibrium Fermi function $ f^0_m=\left[1+{\rm exp}({({\varepsilon_{m\bm k}-\mu)}/{k_BT}}) \right]^{-1}$, where $\mu$ is the chemical potential and $T$ is the temperature. To account for relaxation processes, we apply the adiabatic switching-on approximation~\cite{mandal_24quantum,Varshney_2023,Das_Lahiri_2023,Maneesh_24band}. Using this, we iteratively solve the quantum Liouville equation to obtain the second-order density matrix, $\rho^{(2)}({\bm k},t)$. 

The second-order nonlinear spin magnetization is then evaluated using $\delta {S}^{(2)}_a (t)= \int_{\bm k} \, {\rm Tr}[\hat{s}^a \hat{\rho}^{(2)}({\bm k},t)]=\sum_{mp} \int_{\bm k} \rho_{mp}^{(2)}({\bm k},t) s^a_{pm}$. Here, $s^a_{pm}=\bra{u_{p \bm k}} \hat{s}^a \ket{u_{m \bm k}}$  represents the expectation value of the spin operator $\hat{s}$ along the $a$-th direction, and $\int_{\bm k} \equiv \int d^dk/{(2\pi)^d}$ is the integration over momentum space in $d$ dimensions. The resulting LNSM, which is second order in the electric field, can be expressed in terms of the spin susceptibility tensor $\alpha_{a;bc}(\omega,\mu)$ as: 
\be
  \delta S^{(2)}_a(\omega,\mu) = 
 \left[{\alpha}_{a;bc}^{\rm shg} E_b E_c e^{-i2\omega t} + {\rm c.c.}\right] + \alpha_{a;bc}^{\rm rec}E_{b}^{*} E_c~.  \label{delta_S2}
\ee
The superscript `shg' and `rec' refer to the second harmonic generation and rectification contributions, respectively. We present the detailed calculations of the linear and nonlinear spin susceptibility in Sec.~S1 of the Supplemental Material (SM)~\footnote{The Supplemental Material discuss: S1) Detailed calculation of light-induced spin magnetization; S2) Gauge invariance of spin susceptibilities, S3) Spin susceptibilities under different light polarizations; S4) Angular dependence of spin magnetization under linearly polarized light; S5) Detailed symmetry analysis; S6) Semimetallic antiferromagnet: quasi-2D model of {\ch{CuMnAs}}; S7) Applications of nonlinear spin magnetization.}.

We find that the $\bm k$-resolved spin susceptibility components, 
${\alpha}_{a;bc}(\omega,\mu) =  \int_{\bm k} \tilde{\alpha}_{a;bc}(\omega,\bm k)$, can be decomposed into the following contributions: 
\be \label{alpha_tilde}
\tilde{\alpha}_{a;bc}=\tilde{\alpha}_{a;bc}^{\rm D} + \tilde{\alpha}_{a;bc}^{\rm ASP} + \tilde{\alpha}_{a;bc}^{\rm SSh}+\tilde{\alpha}_{a;bc}^{\rm SI}+\tilde{\alpha}_{a;bc}^{\rm VI}+\tilde{\alpha}_{a;bc}^{\rm MB,1}+\tilde{\alpha}_{a;bc}^{\rm MB,2}.
\ee 
We summarize the functional forms of these components for the second harmonic and rectification contribution to LNSM in Table~\ref{Table_1} and Table S1  in SM~\cite{Note2}, respectively. {Table~\ref{Table_1} shows that ${\tilde \alpha}_{a;bc}^{\rm D}$ and ${\tilde \alpha}_{a;bc}^{\rm ASM}$ arise from Fermi surface contributions, which are finite only in metallic systems. In contrast, all other contributions stem from the Fermi sea and are finite for metallic and insulating systems. Notably, all the spin susceptibility components presented in Table~\ref{Table_1} are U(1) gauge-invariant, ensuring their applicability for 
use with {\it ab initio} methods (see Sec.~S2 of SM~\cite{Note2} for details). 
%

\begin{table}[t!]
    \begin{center}
    \caption{Band geometric origin of different contributions to the $\bm k$-resolved second harmonic nonlinear spin susceptibilities. We express the susceptibilities as ${\tilde \alpha_{a;bc}} =  \sum \rm (Ocu.~Fn. \times JDOS \times Band~Geom.)$, where the summation includes all relevant band indices. Here, `Ocu. Fn.', `JDOS', and `Band Geom.' represent the occupation function ($f_m^0$), the joint density of states term, and band-geometric quantities, respectively. We denote $\omega_{mp}=(\varepsilon_{m\bm k}-\varepsilon_{p\bm k})/\hbar$,  $\hbar v_{mm}^b = \partial_{b} \varepsilon_{m\bm k}$, $g_{0}^{N\omega}= \left[1/\tau-iN\omega \right]^{-1}$, and $g_{mp}^{N\omega}=\left[ N/\tau-i(N\omega-\omega_{mp})\right]^{-1}$ where $\partial_b \equiv \partial_{k_b}$. We have defined $F_{mp}=f_m^0-f_p^0$, and the covariant derivative $D_{mp}^b=\partial_b-i(\mathcal{R}_{mm}^b-\mathcal{R}_{pp}^b)$ with $\mathcal{R}_{mp}^a=\bra{u_{m \bm k}}i\partial_{a} \ket{u_{p \bm k}}$ being the Berry connection.} 
    \setlength\tabcolsep{0.18cm}
    \renewcommand{\arraystretch}{1.6}
    \begin{tabular}{c c c c c }
    \hline
    \hline 
   $\tilde{\alpha}_{a;bc}^{\rm shg}$ & Ocu. Fn. & JDOS &  Band Geom.  \\
    \hline
    \hline
    
    $\tilde{\alpha}_{a;bc}^{\rm D}$ & $ \partial_b\partial_c f_m^0$ & $g_0^\omega g_0^{2\omega} $ & $s_{mm}^a$  \\

    {$\tilde{\alpha}_{a;bc}^{\rm ASM}$} & $\partial_b F_{mp}$ & $g_{mp}^{2\omega} (g_0^\omega +g_{mp}^\omega) $ &  $  i s_{pm}^a \mathcal{R}_{mp}^c$\\

    {$\tilde{\alpha}_{a;bc}^{\rm SSh}$} & $F_{mp}$ & $g_{mp}^{2\omega} g_{mp}^\omega$ & $ i s_{pm}^a D_{mp}^b \mathcal{R}_{mp}^c$ \\

    {$\tilde{\alpha}_{a;bc}^{\rm VI}$}  & $F_{mp}$ &   $\omega_{mp} g_{mp}^{2\omega} \frac{\partial g_{mp}^{\omega}}{\partial \omega_{mp}}$ & $ i s_{pm}^a \dfrac{(v_{mm}^b-v_{pp}^b)}{\omega_{mp}} \mathcal{R}_{mp}^c$\\

    {$\tilde{\alpha}_{a;bc}^{\rm SI}$}  & $F_{mp}$ & $ g_0^{2\omega} g_{mp}^\omega$ & $ (s_{pp}^a - s_{mm}^a) \mathcal{R}_{mp}^{b} \mathcal{R}_{pm}^{c}$ \\

    {$\tilde{\alpha}_{a;bc}^{\rm MB,1}$} & $F_{np}$ & $ g_{mp}^{2\omega} g_{np}^{\omega}$ & $s_{pm}^a \mathcal{R}_{mn}^b\mathcal{R}_{np}^c$  \\

    {$\tilde{\alpha}_{a;bc}^{\rm MB,2}$} & $F_{np}$ & $ g_{pm}^{2\omega} g_{pn}^{\omega}$ & $ s_{mp}^a \mathcal{R}_{nm}^b\mathcal{R}_{pn}^c$\\
    
    \hline
    \hline
    \end{tabular}
    \label{Table_1}
    \end{center}
\end{table}

\textcolor{blue}{\it {Band geometric origin of LNSM:--}} 
We now highlight the band-geometric origin of the different physical processes generating LNSM in Eq.~\eqref{alpha_tilde} and Table~\ref{Table_1}. \\
i){\it Drude Contribution}-${\tilde{\alpha}}^{\rm D}_{a;bc}$: This is an intraband response, which arises from the electric field-induced shift in the Fermi surface. It is proportional to the spin quadrupole moment~\cite{Fregoso_2022}, expressed as $\int_{\bm k} s_{mm}^a \partial_b \partial_c f_m^0 \propto \int_{\bm k} f_m^0 \partial_b \partial_c s_{mm}^a$. \\
ii) {\it Anomalous Spin Magnetization (ASM)}-${\tilde \alpha}^{\rm ASM}_{a;bc}$: This term reflects the generation of an anomalous spin dipole moment, $\propto s_{pm}^a {\cal R}^c_{mp} \partial_b f_m^0$ in response to an electric field~\cite{Xiao_2023}. The ASM mechanism for LNSM is analogous to the Berry curvature dipole-induced nonlinear anomalous Hall effect~\cite{Sodemann_prl15}. \\
iii) {\it Spin Shift (SSh)}-${\tilde \alpha}^{\rm SSh}_{a;bc}$: The spin density change during the shift of wave packets in real space gives rise to $\tilde{\alpha}_{a;bc}^{\rm SSh}$~\footnote{This is because $\tilde{\alpha}_{a;bc}^{\rm SSh}$ contains shift-vector ($A^{bc}_{mp}$) in the form of $D_{mp}^b \mathcal{R}_{mp}^c=-i{\cal R}_{mp}^c A^{bc}_{mp}$~\cite{Sipe_prb00,Bhalla_2022}, which signifies the position shift of real-space wavefunction of Bloch electrons as shown in Fig.~\ref{Fig_1}(b).}. ${\tilde{\alpha}}^{\rm SSh}_{a;bc}$ is similar to the position shift induced shift conductivity in the photogalvanic effect~\cite{Sipe_prb00}.
\\
iv)  {\it Velocity Injection (VI)}-${\tilde \alpha}^{\rm VI}_{a;bc}$: Due to the difference in band velocities ($v_{mm}^b - v_{pp}^b$) of optically excited electrons along the direction of the applied field, ${\tilde \alpha}^{\rm VI}_{a;bc}$ is generated. It is analogous to the velocity injection-based second-order nonlinear charge currents in the photovoltaic effect~\cite{Aversa_1995,Sipe_prb00,Bhalla_2022}. \\ 
v) {\it Spin Injection (SI)}-${\tilde \alpha}^{\rm SI}_{a;bc}$: Spin magnetization can also be driven by interband spin injection. It arises from the changes in spin angular momentum, $\propto (s^a_{pp} - s^a_{mm})$, during optical excitation across bands.\\
vi) {\it Multi-Band (MB)}-${\tilde \alpha}^{\rm MB,1/2}_{a;bc}$: These contributions emerge from transitions involving more than two bands and are significant in systems with multiple bands. 

Figure~\ref{Fig_1} presents a schematic of the key physical mechanisms driving the second harmonic and rectification responses of LNSM. These encompass a wide range of effects, including Fermi sea and Fermi surface contributions and the impact of disorder. Notably, these contributions and their underlying physical processes have not been thoroughly explored in prior studies.

\textcolor{blue}{\it {LNSM under different polarization of light:--}} 
In practical scenarios, it is crucial to translate the calculated LNSM susceptibilities into specific spin susceptibilities based on the polarization of the incident light. To analyze the LNSM for linearly polarized light (LPL, denoted by $\updownarrow$), we consider ${\bm E}(t) = |{\bm E}|(\cos\theta,\sin\theta,0) \{{ e}^{-i\omega t}+ {\rm c.c.} \}$, where $\theta$ is the angle between ${\bm E}(t)$ and $\hat{\bm x}$. We obtain the rectification and second harmonic component of the spin response to be (see Sec.~S3 of SM~\cite{Note2})
\be \label{response_LPL}
\delta S^{\rm rec}_{a,\updownarrow} = \alpha^{\rm rec}_{a,\updownarrow} |{\bm E}|^2~; ~~ \delta S^{\rm shg}_{a,\updownarrow} = \alpha^{\rm shg}_{a,\updownarrow} |{\bm E}|^2 \cos(2\omega t)~. 
\ee 
Here, we denoted
\bea 
\alpha^{\rm rec}_{a,\updownarrow} & = &  
{\rm Re}\left[\alpha_{a;xx}^{\rm rec} \cos^2\theta + \alpha_{a;yy}^{\rm rec} \sin^2\theta + \alpha_{a;xy}^{\rm rec} \sin 2\theta \right], \nonumber \\ 
\alpha^{\rm shg}_{a,\updownarrow} & = & 2 {\rm Re}\left[\alpha_{a;xx}^{\rm shg} \cos^2\theta + \alpha_{a;yy}^{\rm shg} \sin^2\theta + \alpha_{a;xy}^{\rm shg} \sin 2\theta \right]. \nonumber
\eea
The amplitude of the LNSM for LPL is dictated solely by the real part of the corresponding spin susceptibilities and the polarization angle $\theta$ (see also Sec.~S4 of SM~\cite{Note2}).

For the case of circularly polarized light (CPL, denoted by $\circlearrowleft$), we consider 
${\bm E}(t) = |{\bm E}|\{ (1,\sigma i,0) {e}^{-i\omega t} + {\rm c.c.}\}$~, 
with $\sigma=+1~(-1)$ for left (right) circular polarization. 
We obtain the rectification and second harmonic contributions to LNSM to be (see Sec.~S3 of SM~\cite{Note2})
\be  \label{response_CPL}
 \delta S^{\rm rec}_{a,\circlearrowleft} = \alpha^{\rm rec}_{a,\circlearrowleft} |{\bm E}|^2~; 
 ~~\delta S^{\rm shg}_{a,\circlearrowleft} = \alpha^{\rm shg}_{a,\circlearrowleft} |{\bm E}|^2\cos(2\omega t)~,
\ee 
with, 
\bea 
 \alpha^{\rm rec}_{a,\circlearrowleft} &=& {\rm Re} \left[\alpha_{a;xx}^{\rm rec} + \alpha_{a;yy}^{\rm rec}\right]- 2\sigma  \, {\rm Im}[\alpha_{a;xy}^{\rm rec}]~, \nonumber \\ 
 \alpha^{\rm shg}_{a,\circlearrowleft} &=& 2{\rm Re}\left[\alpha_{a;xx}^{\rm shg} - \alpha_{a;yy}^{\rm shg}\right] -4 \sigma \, {\rm Im}[\alpha_{a;xy}^{\rm shg}]~.\nonumber
 \eea
 In contrast to the case for LPL, the CPL-induced nonlinear spin magnetization amplitude also depends on the imaginary part of the spin susceptibilities. Equations~\eqref{response_LPL} and ~\eqref{response_CPL} combined with Table~\ref{Table_1} describe the general expressions for LNSM in both centrosymmetric and non-centrosymmetric materials. These are the central results of this Letter.

 
\begin{table}[t!]
    \begin{center}
    \caption{The symmetry restriction of $\alpha_{a;bc}^{\rm even}$ and $\alpha_{a; bc}^{\rm odd}$ under non-magnetic crystallographic point group elements, considering $\bm E$ lies in the $xy$-plane. 
    The \xmark~(\cmark) indicates that the corresponding response tensor is symmetry forbidden (allowed).} %
    \renewcommand{\arraystretch}{1.5}
    \setlength\tabcolsep{0.18 cm}
    \begin{tabular}{ c | c c c c c c c c c c}
    \hline
    \hline 
    $\alpha_{a;bc}$ &  ${\cal M}_x$ & ${\cal M}_y$ & ${\cal M}_z$ & ${\cal C}_{2,4,6}^x$ & ${\cal C}_{2}^y$ & ${\cal C}_2^z$ &  ${\cal C}_3^x$  & ${\cal C}_{3,4,6}^z$  \\
    \hline
    \hline
    
    $\alpha_{z;xx}$ & \xmark & \xmark & \cmark & \xmark &\xmark & \cmark & \xmark & \cmark  \\
    
    $\alpha_{z;xy}$  & \cmark & \cmark & \cmark & \cmark & \cmark &\cmark &\cmark & \xmark  \\
    
    $\alpha_{z;yy}$  & \xmark & \xmark &  \cmark & \xmark & \xmark & \cmark & \cmark & $\alpha_{z;xx}$    \\
    
    \hline
    \hline
    \end{tabular}
    \label{Table_2}
    \end{center}
\end{table}

\textcolor{blue}{\it Symmetry restrictions on LNSM:--}
To identify crystalline materials that support LNSM, we analyze the impact of discrete crystalline symmetries on spin susceptibilities. Inversion symmetry ($\cal P$) does not impose any restriction on $\alpha_{a;bc}$. Notably, both the real and imaginary parts of the contributions to $\alpha_{a;bc}$, except the Drude term, can have time-reversal ($\cal T$) even and odd components~\footnote{We find that for rectification response of $\alpha_{a;bc}$, only Drude susceptibility is completely $\cal T$-odd, all other susceptibilities have both $\cal T$-even and $\cal T$-odd components. In contrast, for second harmonic response, only the Drude and spin injection is completely $\cal T$-odd. All other components have both $\cal T$ odd and $\cal T$-even counterparts.}. We  denote these by $\alpha_{a;bc}^{\rm even}$ and $\alpha_{a;bc}^{\rm odd}$, respectively (see Table~S3 of SM~\cite{Note2} for detailed expressions). While $\alpha_{a;bc}^{\rm even}$ can be finite in both magnetic and non-magnetic materials, $\alpha_{a;bc}^{\rm odd}$ is finite only in $\cal T$-broken magnetic materials. {This implies, under combined $\cal PT$ symmetry, the $\alpha_{a;bc}^{\rm even}$ ($\alpha_{a;bc}^{\rm odd}$) is nonzero (zero)}.

As third-rank axial tensors, $\alpha_{a;bc}^{\rm even}$ and $\alpha_{a;bc}^{\rm odd}$ obey specific transformation rules under symmetry operations: 
\be
\alpha^{\rm even/odd}_{a';b'c'} = \eta_{\cal T}{\rm det}({\cal O}){\cal O}_{a'a} {\cal O}_{b'b} {\cal O}_{c'c} \alpha^{\rm even/odd}_{a;bc}~.
\ee 
The $\cal O\equiv \cal R$ ($\cal RT$) represents a nonmagnetic (magnetic) point group operation, $\cal R$ being a spatial operation. Here, $\eta_{\cal T}$ depends on the magnetic point group symmetry transformation, with $\eta_{\cal T} = 1$ for $\alpha^{\rm even}_{a;bc}$ under any symmetry operation. For $\alpha^{\rm odd}_{a;bc}$, we have $\eta_{\cal T} = -1$ for magnetic point group operations involving both spatial and time-reversal symmetry, and $\eta_{\cal T}=1$ for non-magnetic point group operations.

For light propagating along the $\hat{\bm z}$-direction, we analyze the restrictions imposed by various non-magnetic (magnetic) point group symmetries on $\alpha_{a;bc}^{\rm even}$ and $\alpha_{a;bc}^{\rm odd}$ in Table~\ref{Table_2} (Table~S4 of the SM~\cite{Note2}). Furthermore, we enumerate all the finite components of the LNSM susceptibility tensor, allowed by crystalline symmetry, for all 122 magnetic point groups in Table~S5 of the SM~\cite{Note2}. Our symmetry analysis highlights the broad range of magnetic and non-magnetic materials that support LNSM. Interestingly, we can induce LNSM in either the in-plane or out-of-plane direction, in a controlled manner depending on the material's symmetry. This control can be leveraged to manipulate local magnetization in magnetic materials. We demonstrate this explicitly for a topological antiferromagnet, CuMnAs.

\textcolor{blue}{\it LNSM in CuMnAs:--}
We consider a minimal model of tetragonal CuMnAs on a crinkled quasi-2D square lattice with a collinear antiferromagnetic state~\cite{smejkal_prl17, Watanabe_prx21}, detailed in Sec.~S6 of SM~\cite{Note2}. The Néel vector lies in the $xy$ plane, as shown in Fig.~\ref{Fig_2}(a). 
CuMnAs hosts the combined $\cal PT$ ${\cal C}_{2}^z$, ${\cal M}_x$, and ${\cal M}_y$ symmetries, while breaking individual $\cal P$ and $\cal T$ symmetries. Although the broken $\cal P$ symmetry allows a linear spin magnetization $\delta S^{(1)}_a = \alpha_{a;b} E_b$, only $\alpha_{x;y}$ and $\alpha_{y;x}$ are nonzero~\cite{Note2}. However, the corresponding $\delta S^{(1)}_x$ and $\delta S^{(1)}_y$ are significantly smaller than $\delta S^{(2)}_a$~\cite{Note2}, making $\delta S^{(2)}_a$ the dominant response in CuMnAs. As shown in Table~\ref{Table_2}, only $\alpha_{z;xy}$ is finite, while all other out-of-plane ($\alpha_{z;xx}$ and $\alpha_{z;yy}$) and in-plane response tensors (see Table~S4 in SM~\cite{Note2}) of LNSM vanish due to the spatial symmetries of CuMnAs.

\begin{figure}[t]
    \centering
    \includegraphics[width=1.0\linewidth]{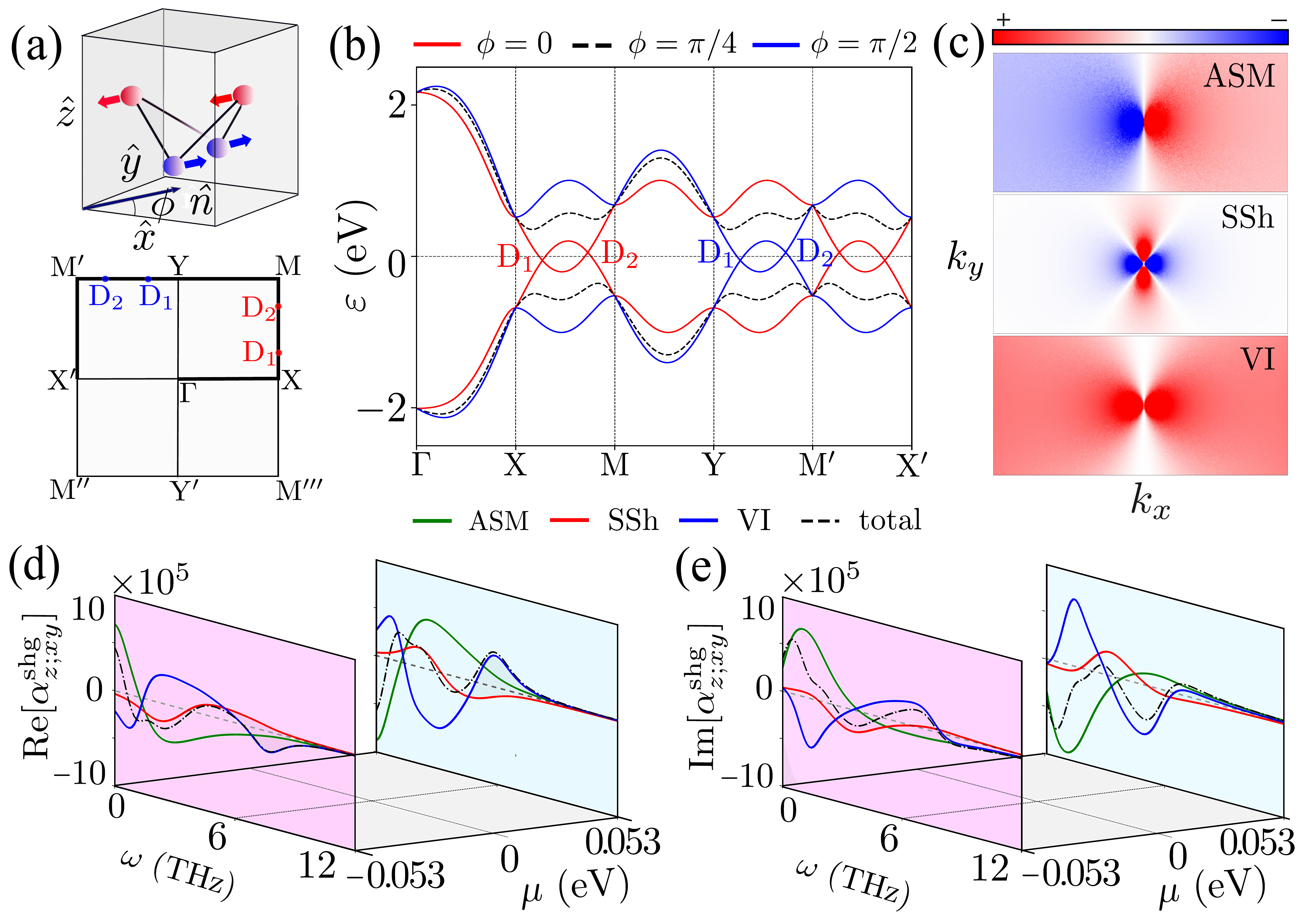}
    \caption{{\bf N\'eel vector dependent LNSM in CuMnAs}. (a) Top: the N\'eel vector orientation in quasi-2D tetragonal CuMnAs. Bottom: The 2D Brillouin zone projection of CuMnAs. (b) The band dispersion of CuMnAs along the $\rm \Gamma$-X-M-Y-$\rm M'$-$\rm X'$ path for three different orientations of in-plane N\'eel vector. (c) $\bm k$-space distribution of the spin susceptibility contributions arising from the anomalous spin magnetization (ASM), spin shift (SSh) and velocity injection (VI) terms for one of the valence bands, near Dirac point (D2) for the N\'eel vector along the $x$-direction ({\it i.e.} $\phi=0)$ and $\omega=3.25$ THz. The nonlinear spin susceptibility components of (d) ${\rm Re}[\alpha^{\rm shg}_{z;xy}]$ and (e) ${\rm Im}[\alpha^{\rm shg}_{z;xy}]$ in units of $\mu_B/\rm V^2$ as a function of light frequency $\omega$ for chemical potential $\mu=-0.053$ eV (D1) and $\mu=0.053$ eV (D2). In (d) and (e), the responses are calculated for $\phi =0$, assuming temperature $T=12$ K, and $\tau=1$ picosecond.}
    \label{Fig_2}
\end{figure}

We investigate the impact of different in-plane N\'eel vector orientations ($\hat{\bm n}$, specified by $\phi$) on the band dispersion of CuMnAs in Fig.~\ref{Fig_2}(b). A semimetal-to-insulator transition occurs as the N\'eel vector rotates within the $xy$ plane. For the semimetallic phase ($\hat{\bm n} = \hat{\bm x}$), we analyze the variation of ${\rm Re}[\alpha_{z;xy}^{\rm shg}]$ and ${\rm Im}[\alpha_{z;xy}^{\rm shg}]$ with $\omega$ at two chemical potentials corresponding to Dirac points D1 and D2, marked by red in Fig.~\ref{Fig_2}(b). The LNSM responses are enhanced near these Dirac points due to significant band geometric contributions in low bandgap materials (see Fig.~\ref{Fig_2}(c) and Fig.~S1 of the SM~\cite{Note2}). We analyze the different spin susceptibility contributions in more detail in Table~S3 of the SM~\cite{Note2}. The presence of $\cal PT$-symmetry forces the Drude and spin injection contributions to vanish. 

\begin{figure}
    \centering
    \includegraphics[width=1.02\linewidth]{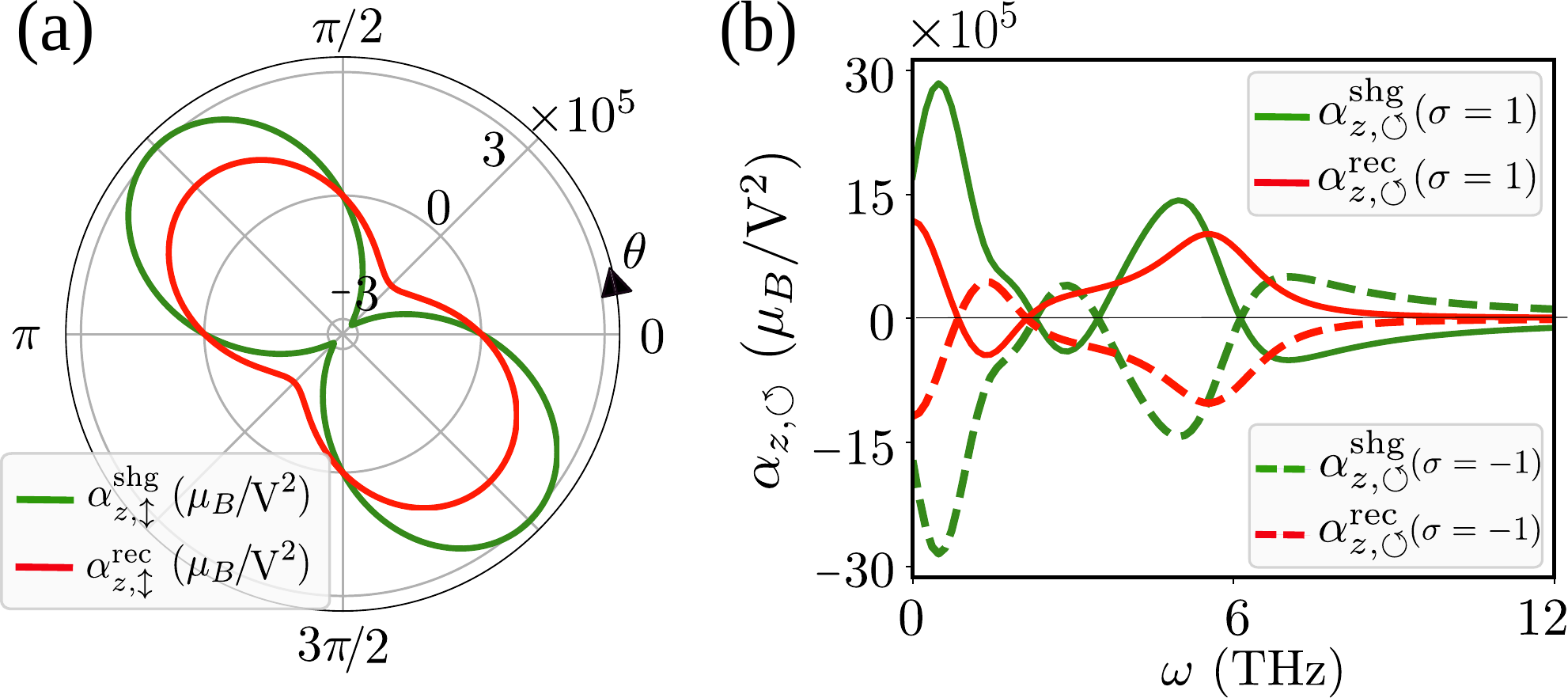}
    \caption{{\bf Polarization control of LNSM}. (a) The $\alpha^{\rm shg}_{z,\updownarrow}$ and $\alpha^{\rm rec}_{z,\updownarrow}$ as a function of the polarization angle $\theta$ for linearly polarized light, at $\omega=3.25$ THz. (b) Light-helicity dependence of the second harmonic and rectification component of total spin susceptibility. The solid (dashed) line represents the responses under the left (right) circularly polarized light, respectively. All the responses are calculated at $\mu=0.053$ eV (D2), $\phi=0$ and $T=12$ K.}
    \label{Fig_3}
\end{figure}

The resonant behavior of $\alpha_{z;xy}^{\rm shg}$ with respect to $\omega$ is evident in Fig.~\ref{Fig_2}(d) and (e), where ${\rm Re}[\alpha_{z;xy}^{\rm shg}]$ and ${\rm Im}[\alpha_{z;xy}^{\rm shg}]$ have maximum value of $5.6\times10^5$ and $7.1\times10^5$ $\mu_B/{\rm V}^2$ at $\omega\sim 0$ and $\omega\sim 0.5$ THz, respectively for $\mu=0.053$ eV. Due to in-plane mirror symmetries, $\alpha_{z;xx}$ and $\alpha_{z;yy}$ are forbidden, making ${\rm Re}[\alpha_{z;xy}^{\rm shg}]$ and ${ \rm Im}[\alpha_{z;xy}^{\rm shg}]$ to be the main contributor to spin magnetization along the $\hat{\bm z}$-axis under LPL and CPL. For an electric field of $E\sim 1$ MV/m (light intensity $\sim 0.27$ MW/$\rm m^2$), the peak spin magnetization reaches $\delta S^{(2)}_z \sim 1.77~(2.22)~ \mu_B/{\rm nm}^3$ under LPL (CPL) illumination~\footnote{Here, we have converted the $\delta S^{(2)}$ in conventional 3D units by dividing the result by the thickness $\sim 6.38$ \AA~\cite{Wadley_sr2015} of the CuMnAs unit cell.}. Note that these values are much larger than the linear spin generation experimentally observed in noncentrosymmetric materials ($\sim 10^{-9}$ to $10^{-6}~\mu_B/{\rm nm^3}$)~\cite{Fang_NN11, Kurebayashi_NN14, Chernyshov_np09}. This highlights that CuMnAs can achieve giant~\footnote{We found the LNSM value of tetragonal CuMnAs in the semimetallic state to be around $1.12~\mu_B/{\rm nm^2}$ under LPL for an electric field amplitude $1$ MV/m. 
We mention that linear-order spin magnetization in ${\rm CuMnAs}$ is typically around $10^{-14}\mu_B/{\rm nm}^2$ for the same field value. In a \ch{CrI3} bilayer, spin magnetization of approximately $0.04\mu_B/{\rm nm}^2$ can be found for the same laser intensity~\cite{Li_prb21}. Similarly, in a single septuple layer of \ch{MnBi2Te4}, the spin magnetization is found to be $0.7\times 10^{-3}~\mu_B/{\rm nm^2}$ for the same electric field value~\cite{Xiao_2023}.} LNSM under THz light irradiation, despite having zero net equilibrium magnetization. Our predicted LNSM can be observed using techniques like magneto-optical Kerr spectroscopy, anisotropic magnetoresistance, or spin-orbit torque measurements~\cite{Satohi_prb24}.

In Fig.~\ref{Fig_3}(a), we show the variation of LNSM responses, $\alpha_{z,\updownarrow}^{\rm shg}$ and $\alpha_{z,\updownarrow}^{\rm rec}$, with the polarization angle $\theta$ of linearly polarized light relative to $\hat{\bm x}$ direction. Both responses exhibit a $\sin 2\theta$ dependence with $\pi$ periodicity (see also Sec.~S4 of SM~\cite{Note2}).
In Fig.~\ref{Fig_3}(b), we show the LNSM responses, $\alpha^{\rm shg}_{z,\circlearrowleft}$ and $\alpha^{\rm rec}_{z,\circlearrowleft}$, for left- and right-circularly polarized light. The sign of induced magnetization changes with CPL helicity, enabling precise control over spin magnetization. This optical control, achievable through both LPL and CPL, is particularly advantageous for light induced nonlinear spin-orbit torque for precise control over magnetization dynamics in magnetic materials. 
We further discuss the LNSM-driven optical spin-orbit torque for magnetization switching and terahertz oscillation generation in Sec.~S7 of SM~\cite{Note2}. 

\textcolor{blue}{\it Conclusions:--} Our prediction of light induced nonlinear spin magnetization significantly expands the potential for optically controlling spin magnetization in both centrosymmetric and non-centrosymmetric materials. This nonlinear mechanism is a powerful tool for probing previously unexplored band geometric properties linked to the spin of Bloch electrons. We show that terahertz light can induce substantial spin magnetization near band crossings in quantum materials, and it offers precise control over magnetic ordering. In addition to advancing the theoretical framework of light-induced spin magnetization, our work lays the foundation for future studies on polarization dependent spin magnetization in centrosymmetric materials.

While our focus has been on the spin magnetization of Bloch electrons, this work also motivates the exploration of light-induced nonlinear orbital magnetization. Investigating nonlinear orbital magnetization and using it to generate orbital torques~\cite{HWLee_prb20,Lee_NatComm2021,Ando_PRR23} could have significant implications. Light-induced spin and orbital magnetization offer new opportunities to control the dynamics of spin and orbital angular momentum, paving the way for innovative opto-spintronic and opto-orbitronic devices.

\textcolor{blue}{\it Acknowledgments:--}
We thank Debasis Dutta and Nirmalya Jana for many illuminating discussions. S.S. acknowledges IIT Kanpur for funding support through the Institute Fellowship. S.D. acknowledges the Ministry of Education, Government of India, for funding support through the Prime Minister's Research Fellowship.


\bibliography{refs}

\begin{thebibliography}{70}%
\makeatletter
\providecommand \@ifxundefined [1]{%
 \@ifx{#1\undefined}
}%
\providecommand \@ifnum [1]{%
 \ifnum #1\expandafter \@firstoftwo
 \else \expandafter \@secondoftwo
 \fi
}%
\providecommand \@ifx [1]{%
 \ifx #1\expandafter \@firstoftwo
 \else \expandafter \@secondoftwo
 \fi
}%
\providecommand \natexlab [1]{#1}%
\providecommand \enquote  [1]{``#1''}%
\providecommand \bibnamefont  [1]{#1}%
\providecommand \bibfnamefont [1]{#1}%
\providecommand \citenamefont [1]{#1}%
\providecommand \href@noop [0]{\@secondoftwo}%
\providecommand \href [0]{\begingroup \@sanitize@url \@href}%
\providecommand \@href[1]{\@@startlink{#1}\@@href}%
\providecommand \@@href[1]{\endgroup#1\@@endlink}%
\providecommand \@sanitize@url [0]{\catcode `\\12\catcode `\$12\catcode
  `\&12\catcode `\#12\catcode `\^12\catcode `\_12\catcode `\%12\relax}%
\providecommand \@@startlink[1]{}%
\providecommand \@@endlink[0]{}%
\providecommand \url  [0]{\begingroup\@sanitize@url \@url }%
\providecommand \@url [1]{\endgroup\@href {#1}{\urlprefix }}%
\providecommand \urlprefix  [0]{URL }%
\providecommand \Eprint [0]{\href }%
\providecommand \doibase [0]{http://dx.doi.org/}%
\providecommand \selectlanguage [0]{\@gobble}%
\providecommand \bibinfo  [0]{\@secondoftwo}%
\providecommand \bibfield  [0]{\@secondoftwo}%
\providecommand \translation [1]{[#1]}%
\providecommand \BibitemOpen [0]{}%
\providecommand \bibitemStop [0]{}%
\providecommand \bibitemNoStop [0]{.\EOS\space}%
\providecommand \EOS [0]{\spacefactor3000\relax}%
\providecommand \BibitemShut  [1]{\csname bibitem#1\endcsname}%
\let\auto@bib@innerbib\@empty
\bibitem [{\citenamefont {Ma}\ \emph {et~al.}(2023)\citenamefont {Ma},
  \citenamefont {Krishna~Kumar}, \citenamefont {Xu}, \citenamefont {Koppens},\
  and\ \citenamefont {Song}}]{Song_nrp23}%
  \BibitemOpen
  \bibfield  {author} {\bibinfo {author} {\bibfnamefont {Qiong}\ \bibnamefont
  {Ma}}, \bibinfo {author} {\bibfnamefont {Roshan}\ \bibnamefont
  {Krishna~Kumar}}, \bibinfo {author} {\bibfnamefont {Su-Yang}\ \bibnamefont
  {Xu}}, \bibinfo {author} {\bibfnamefont {Frank H.~L.}\ \bibnamefont
  {Koppens}}, \ and\ \bibinfo {author} {\bibfnamefont {Justin C.~W.}\
  \bibnamefont {Song}},\ }\bibfield  {title} {\enquote {\bibinfo {title}
  {Photocurrent as a multiphysics diagnostic of quantum materials},}\ }\href
  {\doibase 10.1038/s42254-022-00551-2} {\bibfield  {journal} {\bibinfo
  {journal} {Nature Reviews Physics}\ }\textbf {\bibinfo {volume} {5}},\
  \bibinfo {pages} {170–184} (\bibinfo {year} {2023})}\BibitemShut {NoStop}%
\bibitem [{\citenamefont {Schiffrin}\ \emph {et~al.}(2013)\citenamefont
  {Schiffrin}, \citenamefont {Paasch-Colberg}, \citenamefont {Karpowicz},
  \citenamefont {Apalkov}, \citenamefont {Gerster}, \citenamefont
  {M{\"u}hlbrandt}, \citenamefont {Korbman}, \citenamefont {Reichert},
  \citenamefont {Schultze}, \citenamefont {Holzner}, \citenamefont {Barth},
  \citenamefont {Kienberger}, \citenamefont {Ernstorfer}, \citenamefont
  {Yakovlev}, \citenamefont {Stockman},\ and\ \citenamefont
  {Krausz}}]{Schiffrin_nature2013}%
  \BibitemOpen
  \bibfield  {author} {\bibinfo {author} {\bibfnamefont {Agustin}\ \bibnamefont
  {Schiffrin}}, \bibinfo {author} {\bibfnamefont {Tim}\ \bibnamefont
  {Paasch-Colberg}}, \bibinfo {author} {\bibfnamefont {Nicholas}\ \bibnamefont
  {Karpowicz}}, \bibinfo {author} {\bibfnamefont {Vadym}\ \bibnamefont
  {Apalkov}}, \bibinfo {author} {\bibfnamefont {Daniel}\ \bibnamefont
  {Gerster}}, \bibinfo {author} {\bibfnamefont {Sascha}\ \bibnamefont
  {M{\"u}hlbrandt}}, \bibinfo {author} {\bibfnamefont {Michael}\ \bibnamefont
  {Korbman}}, \bibinfo {author} {\bibfnamefont {Joachim}\ \bibnamefont
  {Reichert}}, \bibinfo {author} {\bibfnamefont {Martin}\ \bibnamefont
  {Schultze}}, \bibinfo {author} {\bibfnamefont {Simon}\ \bibnamefont
  {Holzner}}, \bibinfo {author} {\bibfnamefont {Johannes~V.}\ \bibnamefont
  {Barth}}, \bibinfo {author} {\bibfnamefont {Reinhard}\ \bibnamefont
  {Kienberger}}, \bibinfo {author} {\bibfnamefont {Ralph}\ \bibnamefont
  {Ernstorfer}}, \bibinfo {author} {\bibfnamefont {Vladislav~S.}\ \bibnamefont
  {Yakovlev}}, \bibinfo {author} {\bibfnamefont {Mark~I.}\ \bibnamefont
  {Stockman}}, \ and\ \bibinfo {author} {\bibfnamefont {Ferenc}\ \bibnamefont
  {Krausz}},\ }\bibfield  {title} {\enquote {\bibinfo {title}
  {Optical-field-induced current in dielectrics},}\ }\href {\doibase
  10.1038/nature11567} {\bibfield  {journal} {\bibinfo  {journal} {Nature}\
  }\textbf {\bibinfo {volume} {493}},\ \bibinfo {pages} {70--74} (\bibinfo
  {year} {2013})}\BibitemShut {NoStop}%
\bibitem [{\citenamefont {Reserbat-Plantey}\ \emph {et~al.}(2021)\citenamefont
  {Reserbat-Plantey}, \citenamefont {Epstein}, \citenamefont {Torre},
  \citenamefont {Costa}, \citenamefont {Gonçalves}, \citenamefont {Mortensen},
  \citenamefont {Polini}, \citenamefont {Song}, \citenamefont {Peres},\ and\
  \citenamefont {Koppens}}]{Reserbat_ACS2021}%
  \BibitemOpen
  \bibfield  {author} {\bibinfo {author} {\bibfnamefont {Antoine}\ \bibnamefont
  {Reserbat-Plantey}}, \bibinfo {author} {\bibfnamefont {Itai}\ \bibnamefont
  {Epstein}}, \bibinfo {author} {\bibfnamefont {Iacopo}\ \bibnamefont {Torre}},
  \bibinfo {author} {\bibfnamefont {Antonio~T.}\ \bibnamefont {Costa}},
  \bibinfo {author} {\bibfnamefont {P.~A.~D.}\ \bibnamefont {Gonçalves}},
  \bibinfo {author} {\bibfnamefont {N.~Asger}\ \bibnamefont {Mortensen}},
  \bibinfo {author} {\bibfnamefont {Marco}\ \bibnamefont {Polini}}, \bibinfo
  {author} {\bibfnamefont {Justin C.~W.}\ \bibnamefont {Song}}, \bibinfo
  {author} {\bibfnamefont {Nuno M.~R.}\ \bibnamefont {Peres}}, \ and\ \bibinfo
  {author} {\bibfnamefont {Frank H.~L.}\ \bibnamefont {Koppens}},\ }\bibfield
  {title} {\enquote {\bibinfo {title} {Quantum nanophotonics in two-dimensional
  materials},}\ }\href {\doibase 10.1021/acsphotonics.0c01224} {\bibfield
  {journal} {\bibinfo  {journal} {ACS Photonics}\ }\textbf {\bibinfo {volume}
  {8}},\ \bibinfo {pages} {85--101} (\bibinfo {year} {2021})}\BibitemShut
  {NoStop}%
\bibitem [{\citenamefont {Bader}\ and\ \citenamefont
  {Parkin}(2010)}]{Bader_ARC2010}%
  \BibitemOpen
  \bibfield  {author} {\bibinfo {author} {\bibfnamefont {S.D.}\ \bibnamefont
  {Bader}}\ and\ \bibinfo {author} {\bibfnamefont {S.S.P.}\ \bibnamefont
  {Parkin}},\ }\bibfield  {title} {\enquote {\bibinfo {title} {Spintronics},}\
  }\href {\doibase 10.1146/annurev-conmatphys-070909-104123} {\bibfield
  {journal} {\bibinfo  {journal} {Annual Review of Condensed Matter Physics}\
  }\textbf {\bibinfo {volume} {1}},\ \bibinfo {pages} {71–88} (\bibinfo
  {year} {2010})}\BibitemShut {NoStop}%
\bibitem [{\citenamefont {Liu}\ \emph {et~al.}(2020)\citenamefont {Liu},
  \citenamefont {Xia}, \citenamefont {Xiao}, \citenamefont {García~de Abajo},\
  and\ \citenamefont {Sun}}]{Liu_nature2020}%
  \BibitemOpen
  \bibfield  {author} {\bibinfo {author} {\bibfnamefont {Jing}\ \bibnamefont
  {Liu}}, \bibinfo {author} {\bibfnamefont {Fengnian}\ \bibnamefont {Xia}},
  \bibinfo {author} {\bibfnamefont {Di}~\bibnamefont {Xiao}}, \bibinfo {author}
  {\bibfnamefont {F.~Javier}\ \bibnamefont {García~de Abajo}}, \ and\ \bibinfo
  {author} {\bibfnamefont {Dong}\ \bibnamefont {Sun}},\ }\bibfield  {title}
  {\enquote {\bibinfo {title} {Semimetals for high-performance
  photodetection},}\ }\href {\doibase 10.1038/s41563-020-0715-7} {\bibfield
  {journal} {\bibinfo  {journal} {Nature Materials}\ }\textbf {\bibinfo
  {volume} {19}},\ \bibinfo {pages} {830–837} (\bibinfo {year}
  {2020})}\BibitemShut {NoStop}%
\bibitem [{\citenamefont {Ma}\ \emph {et~al.}(2021)\citenamefont {Ma},
  \citenamefont {Grushin},\ and\ \citenamefont {Burch}}]{Ma_NM21}%
  \BibitemOpen
  \bibfield  {author} {\bibinfo {author} {\bibfnamefont {Qiong}\ \bibnamefont
  {Ma}}, \bibinfo {author} {\bibfnamefont {Adolfo~G.}\ \bibnamefont {Grushin}},
  \ and\ \bibinfo {author} {\bibfnamefont {Kenneth~S.}\ \bibnamefont {Burch}},\
  }\bibfield  {title} {\enquote {\bibinfo {title} {Topology and geometry under
  the nonlinear electromagnetic spotlight},}\ }\href {\doibase
  10.1038/s41563-021-00992-7} {\bibfield  {journal} {\bibinfo  {journal}
  {Nature Materials}\ }\textbf {\bibinfo {volume} {20}},\ \bibinfo {pages}
  {1601--1614} (\bibinfo {year} {2021})}\BibitemShut {NoStop}%
\bibitem [{\citenamefont {Hirohata}\ \emph {et~al.}(2020)\citenamefont
  {Hirohata}, \citenamefont {Yamada}, \citenamefont {Nakatani}, \citenamefont
  {Prejbeanu}, \citenamefont {Diény}, \citenamefont {Pirro},\ and\
  \citenamefont {Hillebrands}}]{HIROHATA20}%
  \BibitemOpen
  \bibfield  {author} {\bibinfo {author} {\bibfnamefont {Atsufumi}\
  \bibnamefont {Hirohata}}, \bibinfo {author} {\bibfnamefont {Keisuke}\
  \bibnamefont {Yamada}}, \bibinfo {author} {\bibfnamefont {Yoshinobu}\
  \bibnamefont {Nakatani}}, \bibinfo {author} {\bibfnamefont {Ioan-Lucian}\
  \bibnamefont {Prejbeanu}}, \bibinfo {author} {\bibfnamefont {Bernard}\
  \bibnamefont {Diény}}, \bibinfo {author} {\bibfnamefont {Philipp}\
  \bibnamefont {Pirro}}, \ and\ \bibinfo {author} {\bibfnamefont {Burkard}\
  \bibnamefont {Hillebrands}},\ }\bibfield  {title} {\enquote {\bibinfo {title}
  {Review on spintronics: Principles and device applications},}\ }\href
  {\doibase https://doi.org/10.1016/j.jmmm.2020.166711} {\bibfield  {journal}
  {\bibinfo  {journal} {Journal of Magnetism and Magnetic Materials}\ }\textbf
  {\bibinfo {volume} {509}},\ \bibinfo {pages} {166711} (\bibinfo {year}
  {2020})}\BibitemShut {NoStop}%
\bibitem [{\citenamefont {Jungwirth}\ \emph {et~al.}(2012)\citenamefont
  {Jungwirth}, \citenamefont {Wunderlich},\ and\ \citenamefont
  {Olejník}}]{Jungwirth_12}%
  \BibitemOpen
  \bibfield  {author} {\bibinfo {author} {\bibfnamefont {Tomas}\ \bibnamefont
  {Jungwirth}}, \bibinfo {author} {\bibfnamefont {J\"{o}rg}\ \bibnamefont
  {Wunderlich}}, \ and\ \bibinfo {author} {\bibfnamefont {Kamil}\ \bibnamefont
  {Olejník}},\ }\bibfield  {title} {\enquote {\bibinfo {title} {Spin hall
  effect devices},}\ }\href {\doibase 10.1038/nmat3279} {\bibfield  {journal}
  {\bibinfo  {journal} {Nature Materials}\ }\textbf {\bibinfo {volume} {11}},\
  \bibinfo {pages} {382–390} (\bibinfo {year} {2012})}\BibitemShut {NoStop}%
\bibitem [{\citenamefont {Fert}\ \emph
  {et~al.}(2024{\natexlab{a}})\citenamefont {Fert}, \citenamefont {Ramesh},
  \citenamefont {Garcia}, \citenamefont {Casanova},\ and\ \citenamefont
  {Bibes}}]{Bibes_RMP24}%
  \BibitemOpen
  \bibfield  {author} {\bibinfo {author} {\bibfnamefont {Albert}\ \bibnamefont
  {Fert}}, \bibinfo {author} {\bibfnamefont {Ramamoorthy}\ \bibnamefont
  {Ramesh}}, \bibinfo {author} {\bibfnamefont {Vincent}\ \bibnamefont
  {Garcia}}, \bibinfo {author} {\bibfnamefont {F\`elix}\ \bibnamefont
  {Casanova}}, \ and\ \bibinfo {author} {\bibfnamefont {Manuel}\ \bibnamefont
  {Bibes}},\ }\bibfield  {title} {\enquote {\bibinfo {title} {Electrical
  control of magnetism by electric field and current-induced torques},}\ }\href
  {\doibase 10.1103/RevModPhys.96.015005} {\bibfield  {journal} {\bibinfo
  {journal} {Reviews of Modern Physics}\ }\textbf {\bibinfo {volume} {96}},\
  \bibinfo {pages} {015005} (\bibinfo {year} {2024}{\natexlab{a}})}\BibitemShut
  {NoStop}%
\bibitem [{\citenamefont {\ifmmode \check{Z}\else
  \v{Z}\fi{}uti\ifmmode~\acute{c}\else \'{c}\fi{}}\ \emph
  {et~al.}(2004)\citenamefont {\ifmmode \check{Z}\else
  \v{Z}\fi{}uti\ifmmode~\acute{c}\else \'{c}\fi{}}, \citenamefont {Fabian},\
  and\ \citenamefont {Das~Sarma}}]{sharma_rmp04}%
  \BibitemOpen
  \bibfield  {author} {\bibinfo {author} {\bibfnamefont {Igor}\ \bibnamefont
  {\ifmmode \check{Z}\else \v{Z}\fi{}uti\ifmmode~\acute{c}\else \'{c}\fi{}}},
  \bibinfo {author} {\bibfnamefont {Jaroslav}\ \bibnamefont {Fabian}}, \ and\
  \bibinfo {author} {\bibfnamefont {S.}~\bibnamefont {Das~Sarma}},\ }\bibfield
  {title} {\enquote {\bibinfo {title} {Spintronics: Fundamentals and
  applications},}\ }\href {\doibase 10.1103/RevModPhys.76.323} {\bibfield
  {journal} {\bibinfo  {journal} {Review of Modern Physics}\ }\textbf {\bibinfo
  {volume} {76}},\ \bibinfo {pages} {323--410} (\bibinfo {year}
  {2004})}\BibitemShut {NoStop}%
\bibitem [{\citenamefont {Puebla}\ \emph {et~al.}(2020)\citenamefont {Puebla},
  \citenamefont {Kim}, \citenamefont {Kondou},\ and\ \citenamefont
  {Otani}}]{Puebla_cm20}%
  \BibitemOpen
  \bibfield  {author} {\bibinfo {author} {\bibfnamefont {Jorge}\ \bibnamefont
  {Puebla}}, \bibinfo {author} {\bibfnamefont {Junyeon}\ \bibnamefont {Kim}},
  \bibinfo {author} {\bibfnamefont {Kouta}\ \bibnamefont {Kondou}}, \ and\
  \bibinfo {author} {\bibfnamefont {Yoshichika}\ \bibnamefont {Otani}},\
  }\bibfield  {title} {\enquote {\bibinfo {title} {Spintronic devices for
  energy-efficient data storage and energy harvesting},}\ }\href {\doibase
  10.1038/s43246-020-0022-5} {\bibfield  {journal} {\bibinfo  {journal}
  {Communications Materials}\ }\textbf {\bibinfo {volume} {1}},\ \bibinfo
  {pages} {24} (\bibinfo {year} {2020})}\BibitemShut {NoStop}%
\bibitem [{\citenamefont {Manchon}\ \emph {et~al.}(2019)\citenamefont
  {Manchon}, \citenamefont {\ifmmode~\check{Z}\else \v{Z}\fi{}elezn\'y},
  \citenamefont {Miron}, \citenamefont {Jungwirth}, \citenamefont {Sinova},
  \citenamefont {Thiaville}, \citenamefont {Garello},\ and\ \citenamefont
  {Gambardella}}]{manchon_rmp19}%
  \BibitemOpen
  \bibfield  {author} {\bibinfo {author} {\bibfnamefont {A.}~\bibnamefont
  {Manchon}}, \bibinfo {author} {\bibfnamefont {J.}~\bibnamefont
  {\ifmmode~\check{Z}\else \v{Z}\fi{}elezn\'y}}, \bibinfo {author}
  {\bibfnamefont {I.~M.}\ \bibnamefont {Miron}}, \bibinfo {author}
  {\bibfnamefont {T.}~\bibnamefont {Jungwirth}}, \bibinfo {author}
  {\bibfnamefont {J.}~\bibnamefont {Sinova}}, \bibinfo {author} {\bibfnamefont
  {A.}~\bibnamefont {Thiaville}}, \bibinfo {author} {\bibfnamefont
  {K.}~\bibnamefont {Garello}}, \ and\ \bibinfo {author} {\bibfnamefont
  {P.}~\bibnamefont {Gambardella}},\ }\bibfield  {title} {\enquote {\bibinfo
  {title} {Current-induced spin-orbit torques in ferromagnetic and
  antiferromagnetic systems},}\ }\href {\doibase 10.1103/RevModPhys.91.035004}
  {\bibfield  {journal} {\bibinfo  {journal} {Review of Modern Physics}\
  }\textbf {\bibinfo {volume} {91}},\ \bibinfo {pages} {035004} (\bibinfo
  {year} {2019})}\BibitemShut {NoStop}%
\bibitem [{\citenamefont {Sierra}\ \emph {et~al.}(2021)\citenamefont {Sierra},
  \citenamefont {Fabian}, \citenamefont {Kawakami}, \citenamefont {Roche},\
  and\ \citenamefont {Valenzuela}}]{Sierra_21}%
  \BibitemOpen
  \bibfield  {author} {\bibinfo {author} {\bibfnamefont {Juan~F.}\ \bibnamefont
  {Sierra}}, \bibinfo {author} {\bibfnamefont {Jaroslav}\ \bibnamefont
  {Fabian}}, \bibinfo {author} {\bibfnamefont {Roland~K.}\ \bibnamefont
  {Kawakami}}, \bibinfo {author} {\bibfnamefont {Stephan}\ \bibnamefont
  {Roche}}, \ and\ \bibinfo {author} {\bibfnamefont {Sergio~O.}\ \bibnamefont
  {Valenzuela}},\ }\bibfield  {title} {\enquote {\bibinfo {title} {Van der
  waals heterostructures for spintronics and opto-spintronics},}\ }\href
  {\doibase 10.1038/s41565-021-00936-x} {\bibfield  {journal} {\bibinfo
  {journal} {Nature Nanotechnology}\ }\textbf {\bibinfo {volume} {16}},\
  \bibinfo {pages} {856--868} (\bibinfo {year} {2021})}\BibitemShut {NoStop}%
\bibitem [{\citenamefont {Liu}\ \emph {et~al.}(2023)\citenamefont {Liu},
  \citenamefont {Malik}, \citenamefont {Zhang},\ and\ \citenamefont
  {Yu}}]{Ting_am23}%
  \BibitemOpen
  \bibfield  {author} {\bibinfo {author} {\bibfnamefont {Sheng}\ \bibnamefont
  {Liu}}, \bibinfo {author} {\bibfnamefont {Iftikhar~Ahmed}\ \bibnamefont
  {Malik}}, \bibinfo {author} {\bibfnamefont {Vanessa~Li}\ \bibnamefont
  {Zhang}}, \ and\ \bibinfo {author} {\bibfnamefont {Ting}\ \bibnamefont
  {Yu}},\ }\bibfield  {title} {\enquote {\bibinfo {title} {Lightning the spin:
  Harnessing the potential of 2d magnets in opto-spintronics},}\ }\href
  {\doibase https://doi.org/10.1002/adma.202306920} {\bibfield  {journal}
  {\bibinfo  {journal} {Advanced Materials}\ ,\ \bibinfo {pages} {2306920}}
  (\bibinfo {year} {2023})}\BibitemShut {NoStop}%
\bibitem [{\citenamefont {Fert}\ \emph
  {et~al.}(2024{\natexlab{b}})\citenamefont {Fert}, \citenamefont {Ramesh},
  \citenamefont {Garcia}, \citenamefont {Casanova},\ and\ \citenamefont
  {Bibes}}]{Manuel_rmp24}%
  \BibitemOpen
  \bibfield  {author} {\bibinfo {author} {\bibfnamefont {Albert}\ \bibnamefont
  {Fert}}, \bibinfo {author} {\bibfnamefont {Ramamoorthy}\ \bibnamefont
  {Ramesh}}, \bibinfo {author} {\bibfnamefont {Vincent}\ \bibnamefont
  {Garcia}}, \bibinfo {author} {\bibfnamefont {F\`elix}\ \bibnamefont
  {Casanova}}, \ and\ \bibinfo {author} {\bibfnamefont {Manuel}\ \bibnamefont
  {Bibes}},\ }\bibfield  {title} {\enquote {\bibinfo {title} {Electrical
  control of magnetism by electric field and current-induced torques},}\ }\href
  {\doibase 10.1103/RevModPhys.96.015005} {\bibfield  {journal} {\bibinfo
  {journal} {Review of Modern Physics}\ }\textbf {\bibinfo {volume} {96}},\
  \bibinfo {pages} {015005} (\bibinfo {year} {2024}{\natexlab{b}})}\BibitemShut
  {NoStop}%
\bibitem [{\citenamefont {Edelstein}(1990)}]{EDELSTEIN_90}%
  \BibitemOpen
  \bibfield  {author} {\bibinfo {author} {\bibfnamefont {V.M.}\ \bibnamefont
  {Edelstein}},\ }\bibfield  {title} {\enquote {\bibinfo {title} {Spin
  polarization of conduction electrons induced by electric current in
  two-dimensional asymmetric electron systems},}\ }\href {\doibase
  https://doi.org/10.1016/0038-1098(90)90963-C} {\bibfield  {journal} {\bibinfo
   {journal} {Solid State Communications}\ }\textbf {\bibinfo {volume} {73}},\
  \bibinfo {pages} {233--235} (\bibinfo {year} {1990})}\BibitemShut {NoStop}%
\bibitem [{\citenamefont {\ifmmode~\check{Z}\else \v{Z}\fi{}elezn\'y}\ \emph
  {et~al.}(2017{\natexlab{a}})\citenamefont {\ifmmode~\check{Z}\else
  \v{Z}\fi{}elezn\'y}, \citenamefont {Gao}, \citenamefont {Manchon},
  \citenamefont {Freimuth}, \citenamefont {Mokrousov}, \citenamefont {Zemen},
  \citenamefont {Ma\ifmmode~\check{s}\else \v{s}\fi{}ek}, \citenamefont
  {Sinova},\ and\ \citenamefont {Jungwirth}}]{Zelenzy_prb17}%
  \BibitemOpen
  \bibfield  {author} {\bibinfo {author} {\bibfnamefont {J.}~\bibnamefont
  {\ifmmode~\check{Z}\else \v{Z}\fi{}elezn\'y}}, \bibinfo {author}
  {\bibfnamefont {H.}~\bibnamefont {Gao}}, \bibinfo {author} {\bibfnamefont
  {Aur\'elien}\ \bibnamefont {Manchon}}, \bibinfo {author} {\bibfnamefont
  {Frank}\ \bibnamefont {Freimuth}}, \bibinfo {author} {\bibfnamefont {Yuriy}\
  \bibnamefont {Mokrousov}}, \bibinfo {author} {\bibfnamefont {J.}~\bibnamefont
  {Zemen}}, \bibinfo {author} {\bibfnamefont {J.}~\bibnamefont
  {Ma\ifmmode~\check{s}\else \v{s}\fi{}ek}}, \bibinfo {author} {\bibfnamefont
  {Jairo}\ \bibnamefont {Sinova}}, \ and\ \bibinfo {author} {\bibfnamefont
  {T.}~\bibnamefont {Jungwirth}},\ }\bibfield  {title} {\enquote {\bibinfo
  {title} {Spin-orbit torques in locally and globally noncentrosymmetric
  crystals: Antiferromagnets and ferromagnets},}\ }\href {\doibase
  10.1103/PhysRevB.95.014403} {\bibfield  {journal} {\bibinfo  {journal}
  {Physical Review B}\ }\textbf {\bibinfo {volume} {95}},\ \bibinfo {pages}
  {014403} (\bibinfo {year} {2017}{\natexlab{a}})}\BibitemShut {NoStop}%
\bibitem [{\citenamefont {S{\'a}nchez}\ \emph {et~al.}(2013)\citenamefont
  {S{\'a}nchez}, \citenamefont {Vila}, \citenamefont {Desfonds}, \citenamefont
  {Gambarelli}, \citenamefont {Attan{\'e}}, \citenamefont {De~Teresa},
  \citenamefont {Mag{\'e}n},\ and\ \citenamefont {Fert}}]{Sanchez_nc13}%
  \BibitemOpen
  \bibfield  {author} {\bibinfo {author} {\bibfnamefont {J.~C.~Rojas}\
  \bibnamefont {S{\'a}nchez}}, \bibinfo {author} {\bibfnamefont
  {L.}~\bibnamefont {Vila}}, \bibinfo {author} {\bibfnamefont {G.}~\bibnamefont
  {Desfonds}}, \bibinfo {author} {\bibfnamefont {S.}~\bibnamefont
  {Gambarelli}}, \bibinfo {author} {\bibfnamefont {J.~P.}\ \bibnamefont
  {Attan{\'e}}}, \bibinfo {author} {\bibfnamefont {J.~M.}\ \bibnamefont
  {De~Teresa}}, \bibinfo {author} {\bibfnamefont {C.}~\bibnamefont
  {Mag{\'e}n}}, \ and\ \bibinfo {author} {\bibfnamefont {A.}~\bibnamefont
  {Fert}},\ }\bibfield  {title} {\enquote {\bibinfo {title} {Spin-to-charge
  conversion using rashba coupling at the interface between non-magnetic
  materials},}\ }\href {\doibase 10.1038/ncomms3944} {\bibfield  {journal}
  {\bibinfo  {journal} {Nature Communications}\ }\textbf {\bibinfo {volume}
  {4}},\ \bibinfo {pages} {2944} (\bibinfo {year} {2013})}\BibitemShut
  {NoStop}%
\bibitem [{\citenamefont {Mellnik}\ \emph {et~al.}(2014)\citenamefont
  {Mellnik}, \citenamefont {Lee}, \citenamefont {Richardella}, \citenamefont
  {Grab}, \citenamefont {Mintun}, \citenamefont {Fischer}, \citenamefont
  {Vaezi}, \citenamefont {Manchon}, \citenamefont {Kim}, \citenamefont
  {Samarth},\ and\ \citenamefont {Ralph}}]{Mellnik_Nat14}%
  \BibitemOpen
  \bibfield  {author} {\bibinfo {author} {\bibfnamefont {A.~R.}\ \bibnamefont
  {Mellnik}}, \bibinfo {author} {\bibfnamefont {J.~S.}\ \bibnamefont {Lee}},
  \bibinfo {author} {\bibfnamefont {A.}~\bibnamefont {Richardella}}, \bibinfo
  {author} {\bibfnamefont {J.~L.}\ \bibnamefont {Grab}}, \bibinfo {author}
  {\bibfnamefont {P.~J.}\ \bibnamefont {Mintun}}, \bibinfo {author}
  {\bibfnamefont {M.~H.}\ \bibnamefont {Fischer}}, \bibinfo {author}
  {\bibfnamefont {A.}~\bibnamefont {Vaezi}}, \bibinfo {author} {\bibfnamefont
  {A.}~\bibnamefont {Manchon}}, \bibinfo {author} {\bibfnamefont {E.-A.}\
  \bibnamefont {Kim}}, \bibinfo {author} {\bibfnamefont {N.}~\bibnamefont
  {Samarth}}, \ and\ \bibinfo {author} {\bibfnamefont {D.~C.}\ \bibnamefont
  {Ralph}},\ }\bibfield  {title} {\enquote {\bibinfo {title} {Spin-transfer
  torque generated by a topological insulator},}\ }\href {\doibase
  10.1038/nature13534} {\bibfield  {journal} {\bibinfo  {journal} {Nature}\
  }\textbf {\bibinfo {volume} {511}},\ \bibinfo {pages} {449--451} (\bibinfo
  {year} {2014})}\BibitemShut {NoStop}%
\bibitem [{\citenamefont {Li}\ \emph {et~al.}(2020)\citenamefont {Li},
  \citenamefont {Chen},\ and\ \citenamefont {Niu}}]{Nui_pnas20}%
  \BibitemOpen
  \bibfield  {author} {\bibinfo {author} {\bibfnamefont {Xiao}\ \bibnamefont
  {Li}}, \bibinfo {author} {\bibfnamefont {Hua}\ \bibnamefont {Chen}}, \ and\
  \bibinfo {author} {\bibfnamefont {Qian}\ \bibnamefont {Niu}},\ }\bibfield
  {title} {\enquote {\bibinfo {title} {Out-of-plane carrier spin in
  transition-metal dichalcogenides under electric current},}\ }\href {\doibase
  10.1073/pnas.1912472117} {\bibfield  {journal} {\bibinfo  {journal}
  {Proceedings of the National Academy of Sciences}\ }\textbf {\bibinfo
  {volume} {117}},\ \bibinfo {pages} {16749--16755} (\bibinfo {year}
  {2020})}\BibitemShut {NoStop}%
\bibitem [{\citenamefont {Johansson}(2024)}]{Annika_pubmed24}%
  \BibitemOpen
  \bibfield  {author} {\bibinfo {author} {\bibfnamefont {Annika}\ \bibnamefont
  {Johansson}},\ }\bibfield  {title} {\enquote {\bibinfo {title} {Theory of
  spin and orbital edelstein effects},}\ }\href {\doibase
  10.1088/1361-648x/ad5e2b} {\bibfield  {journal} {\bibinfo  {journal} {Journal
  of Physics: Condensed Matter}\ }\textbf {\bibinfo {volume} {36}},\ \bibinfo
  {pages} {423002} (\bibinfo {year} {2024})}\BibitemShut {NoStop}%
\bibitem [{\citenamefont {Huang}\ \emph {et~al.}(2024)\citenamefont {Huang},
  \citenamefont {Cao}, \citenamefont {Qiu}, \citenamefont {Bai}, \citenamefont
  {Liao}, \citenamefont {Chen}, \citenamefont {Han}, \citenamefont {Pan},
  \citenamefont {Jin},\ and\ \citenamefont {Song}}]{Song_NC24}%
  \BibitemOpen
  \bibfield  {author} {\bibinfo {author} {\bibfnamefont {Lin}\ \bibnamefont
  {Huang}}, \bibinfo {author} {\bibfnamefont {Yanzhang}\ \bibnamefont {Cao}},
  \bibinfo {author} {\bibfnamefont {Hongsong}\ \bibnamefont {Qiu}}, \bibinfo
  {author} {\bibfnamefont {Hua}\ \bibnamefont {Bai}}, \bibinfo {author}
  {\bibfnamefont {Liyang}\ \bibnamefont {Liao}}, \bibinfo {author}
  {\bibfnamefont {Chong}\ \bibnamefont {Chen}}, \bibinfo {author}
  {\bibfnamefont {Lei}\ \bibnamefont {Han}}, \bibinfo {author} {\bibfnamefont
  {Feng}\ \bibnamefont {Pan}}, \bibinfo {author} {\bibfnamefont {Biaobing}\
  \bibnamefont {Jin}}, \ and\ \bibinfo {author} {\bibfnamefont {Cheng}\
  \bibnamefont {Song}},\ }\bibfield  {title} {\enquote {\bibinfo {title}
  {Terahertz oscillation driven by optical spin-orbit torque},}\ }\href
  {\doibase 10.1038/s41467-024-51440-4} {\bibfield  {journal} {\bibinfo
  {journal} {Nature Communications}\ }\textbf {\bibinfo {volume} {15}},\
  \bibinfo {pages} {7227} (\bibinfo {year} {2024})}\BibitemShut {NoStop}%
\bibitem [{\citenamefont {\ifmmode~\check{Z}\else \v{Z}\fi{}elezn\'y}\ \emph
  {et~al.}(2014{\natexlab{a}})\citenamefont {\ifmmode~\check{Z}\else
  \v{Z}\fi{}elezn\'y}, \citenamefont {Gao}, \citenamefont {V\'yborn\'y},
  \citenamefont {Zemen}, \citenamefont {Ma\ifmmode~\check{s}\else
  \v{s}\fi{}ek}, \citenamefont {Manchon}, \citenamefont {Wunderlich},
  \citenamefont {Sinova},\ and\ \citenamefont {Jungwirth}}]{Zelezny_prl14}%
  \BibitemOpen
  \bibfield  {author} {\bibinfo {author} {\bibfnamefont {J.}~\bibnamefont
  {\ifmmode~\check{Z}\else \v{Z}\fi{}elezn\'y}}, \bibinfo {author}
  {\bibfnamefont {H.}~\bibnamefont {Gao}}, \bibinfo {author} {\bibfnamefont
  {K.}~\bibnamefont {V\'yborn\'y}}, \bibinfo {author} {\bibfnamefont
  {J.}~\bibnamefont {Zemen}}, \bibinfo {author} {\bibfnamefont
  {J.}~\bibnamefont {Ma\ifmmode~\check{s}\else \v{s}\fi{}ek}}, \bibinfo
  {author} {\bibfnamefont {Aur\'elien}\ \bibnamefont {Manchon}}, \bibinfo
  {author} {\bibfnamefont {J.}~\bibnamefont {Wunderlich}}, \bibinfo {author}
  {\bibfnamefont {Jairo}\ \bibnamefont {Sinova}}, \ and\ \bibinfo {author}
  {\bibfnamefont {T.}~\bibnamefont {Jungwirth}},\ }\bibfield  {title} {\enquote
  {\bibinfo {title} {Relativistic n\'eel-order fields induced by electrical
  current in antiferromagnets},}\ }\href {\doibase
  10.1103/PhysRevLett.113.157201} {\bibfield  {journal} {\bibinfo  {journal}
  {Physical Review Letters}\ }\textbf {\bibinfo {volume} {113}},\ \bibinfo
  {pages} {157201} (\bibinfo {year} {2014}{\natexlab{a}})}\BibitemShut
  {NoStop}%
\bibitem [{\citenamefont {Furukawa}\ \emph {et~al.}(2021)\citenamefont
  {Furukawa}, \citenamefont {Watanabe}, \citenamefont {Ogasawara},
  \citenamefont {Kobayashi},\ and\ \citenamefont {Itou}}]{Itou_prr21}%
  \BibitemOpen
  \bibfield  {author} {\bibinfo {author} {\bibfnamefont {Tetsuya}\ \bibnamefont
  {Furukawa}}, \bibinfo {author} {\bibfnamefont {Yuta}\ \bibnamefont
  {Watanabe}}, \bibinfo {author} {\bibfnamefont {Naoki}\ \bibnamefont
  {Ogasawara}}, \bibinfo {author} {\bibfnamefont {Kaya}\ \bibnamefont
  {Kobayashi}}, \ and\ \bibinfo {author} {\bibfnamefont {Tetsuaki}\
  \bibnamefont {Itou}},\ }\bibfield  {title} {\enquote {\bibinfo {title}
  {Current-induced magnetization caused by crystal chirality in nonmagnetic
  elemental tellurium},}\ }\href {\doibase 10.1103/PhysRevResearch.3.023111}
  {\bibfield  {journal} {\bibinfo  {journal} {Physical Review Research}\
  }\textbf {\bibinfo {volume} {3}},\ \bibinfo {pages} {023111} (\bibinfo {year}
  {2021})}\BibitemShut {NoStop}%
\bibitem [{\citenamefont {Droghetti}\ and\ \citenamefont
  {Tokatly}(2023)}]{Ilya_prb23}%
  \BibitemOpen
  \bibfield  {author} {\bibinfo {author} {\bibfnamefont {Andrea}\ \bibnamefont
  {Droghetti}}\ and\ \bibinfo {author} {\bibfnamefont {Ilya~V.}\ \bibnamefont
  {Tokatly}},\ }\bibfield  {title} {\enquote {\bibinfo {title} {Current-induced
  spin polarization at metallic surfaces from first principles},}\ }\href
  {\doibase 10.1103/PhysRevB.107.174433} {\bibfield  {journal} {\bibinfo
  {journal} {Physical Review B}\ }\textbf {\bibinfo {volume} {107}},\ \bibinfo
  {pages} {174433} (\bibinfo {year} {2023})}\BibitemShut {NoStop}%
\bibitem [{\citenamefont {Cao}\ \emph {et~al.}(2024)\citenamefont {Cao},
  \citenamefont {Zeng}, \citenamefont {Li}, \citenamefont {Wang}, \citenamefont
  {Yang}, \citenamefont {Yu},\ and\ \citenamefont {Yao}}]{Yao_prbL24}%
  \BibitemOpen
  \bibfield  {author} {\bibinfo {author} {\bibfnamefont {Jin}\ \bibnamefont
  {Cao}}, \bibinfo {author} {\bibfnamefont {Chuanchang}\ \bibnamefont {Zeng}},
  \bibinfo {author} {\bibfnamefont {Xiao-Ping}\ \bibnamefont {Li}}, \bibinfo
  {author} {\bibfnamefont {Maoyuan}\ \bibnamefont {Wang}}, \bibinfo {author}
  {\bibfnamefont {Shengyuan~A.}\ \bibnamefont {Yang}}, \bibinfo {author}
  {\bibfnamefont {Zhi-Ming}\ \bibnamefont {Yu}}, \ and\ \bibinfo {author}
  {\bibfnamefont {Yugui}\ \bibnamefont {Yao}},\ }\bibfield  {title} {\enquote
  {\bibinfo {title} {Low-frequency divergence of circular photomagnetic effect
  in topological semimetals},}\ }\href {\doibase 10.1103/PhysRevB.110.L041114}
  {\bibfield  {journal} {\bibinfo  {journal} {Physical Review B}\ }\textbf
  {\bibinfo {volume} {110}},\ \bibinfo {pages} {L041114} (\bibinfo {year}
  {2024})}\BibitemShut {NoStop}%
\bibitem [{\citenamefont {Feng}\ \emph
  {et~al.}(2024{\natexlab{a}})\citenamefont {Feng}, \citenamefont {Cao},
  \citenamefont {Zhang}, \citenamefont {Ang}, \citenamefont {Lai},
  \citenamefont {Jiang}, \citenamefont {Xiao},\ and\ \citenamefont
  {Yang}}]{Yang_arxiv24}%
  \BibitemOpen
  \bibfield  {author} {\bibinfo {author} {\bibfnamefont {Xukun}\ \bibnamefont
  {Feng}}, \bibinfo {author} {\bibfnamefont {Jin}\ \bibnamefont {Cao}},
  \bibinfo {author} {\bibfnamefont {Zhi-Fan}\ \bibnamefont {Zhang}}, \bibinfo
  {author} {\bibfnamefont {Lay~Kee}\ \bibnamefont {Ang}}, \bibinfo {author}
  {\bibfnamefont {Shen}\ \bibnamefont {Lai}}, \bibinfo {author} {\bibfnamefont
  {Hua}\ \bibnamefont {Jiang}}, \bibinfo {author} {\bibfnamefont {Cong}\
  \bibnamefont {Xiao}}, \ and\ \bibinfo {author} {\bibfnamefont {Shengyuan~A.}\
  \bibnamefont {Yang}},\ }\href {https://arxiv.org/abs/2409.09669} {\enquote
  {\bibinfo {title} {Intrinsic dynamic generation of spin polarization by
  time-varying electric field},}\ } (\bibinfo {year} {2024}{\natexlab{a}}),\
  \Eprint {http://arxiv.org/abs/2409.09669} {arXiv:2409.09669
  [cond-mat.mes-hall]} \BibitemShut {NoStop}%
\bibitem [{\citenamefont {Xiao}\ \emph {et~al.}(2022)\citenamefont {Xiao},
  \citenamefont {Liu}, \citenamefont {Wu}, \citenamefont {Wang}, \citenamefont
  {Niu},\ and\ \citenamefont {Yang}}]{Xiao_2022}%
  \BibitemOpen
  \bibfield  {author} {\bibinfo {author} {\bibfnamefont {Cong}\ \bibnamefont
  {Xiao}}, \bibinfo {author} {\bibfnamefont {Huiying}\ \bibnamefont {Liu}},
  \bibinfo {author} {\bibfnamefont {Weikang}\ \bibnamefont {Wu}}, \bibinfo
  {author} {\bibfnamefont {Hui}\ \bibnamefont {Wang}}, \bibinfo {author}
  {\bibfnamefont {Qian}\ \bibnamefont {Niu}}, \ and\ \bibinfo {author}
  {\bibfnamefont {Shengyuan~A.}\ \bibnamefont {Yang}},\ }\bibfield  {title}
  {\enquote {\bibinfo {title} {Intrinsic nonlinear electric spin generation in
  centrosymmetric magnets},}\ }\href {\doibase 10.1103/PhysRevLett.129.086602}
  {\bibfield  {journal} {\bibinfo  {journal} {Physical Review Letters}\
  }\textbf {\bibinfo {volume} {129}},\ \bibinfo {pages} {086602} (\bibinfo
  {year} {2022})}\BibitemShut {NoStop}%
\bibitem [{\citenamefont {Xiao}\ \emph {et~al.}(2023)\citenamefont {Xiao},
  \citenamefont {Wu}, \citenamefont {Wang}, \citenamefont {Huang},
  \citenamefont {Feng}, \citenamefont {Liu}, \citenamefont {Guo}, \citenamefont
  {Niu},\ and\ \citenamefont {Yang}}]{Xiao_2023}%
  \BibitemOpen
  \bibfield  {author} {\bibinfo {author} {\bibfnamefont {Cong}\ \bibnamefont
  {Xiao}}, \bibinfo {author} {\bibfnamefont {Weikang}\ \bibnamefont {Wu}},
  \bibinfo {author} {\bibfnamefont {Hui}\ \bibnamefont {Wang}}, \bibinfo
  {author} {\bibfnamefont {Yue-Xin}\ \bibnamefont {Huang}}, \bibinfo {author}
  {\bibfnamefont {Xiaolong}\ \bibnamefont {Feng}}, \bibinfo {author}
  {\bibfnamefont {Huiying}\ \bibnamefont {Liu}}, \bibinfo {author}
  {\bibfnamefont {Guang-Yu}\ \bibnamefont {Guo}}, \bibinfo {author}
  {\bibfnamefont {Qian}\ \bibnamefont {Niu}}, \ and\ \bibinfo {author}
  {\bibfnamefont {Shengyuan~A.}\ \bibnamefont {Yang}},\ }\bibfield  {title}
  {\enquote {\bibinfo {title} {Time-reversal-even nonlinear current induced
  spin polarization},}\ }\href {\doibase 10.1103/PhysRevLett.130.166302}
  {\bibfield  {journal} {\bibinfo  {journal} {Physical Review Letters}\
  }\textbf {\bibinfo {volume} {130}},\ \bibinfo {pages} {166302} (\bibinfo
  {year} {2023})}\BibitemShut {NoStop}%
\bibitem [{\citenamefont {Fregoso}(2022)}]{Fregoso_2022}%
  \BibitemOpen
  \bibfield  {author} {\bibinfo {author} {\bibfnamefont {Benjamin~M.}\
  \bibnamefont {Fregoso}},\ }\bibfield  {title} {\enquote {\bibinfo {title}
  {Bulk photospin effect: Calculation of electric spin susceptibility to second
  order in an electric field},}\ }\href {\doibase 10.1103/PhysRevB.106.195108}
  {\bibfield  {journal} {\bibinfo  {journal} {Physical Review B}\ }\textbf
  {\bibinfo {volume} {106}},\ \bibinfo {pages} {195108} (\bibinfo {year}
  {2022})}\BibitemShut {NoStop}%
\bibitem [{\citenamefont {Zhou}(2022)}]{Zhou_nature22}%
  \BibitemOpen
  \bibfield  {author} {\bibinfo {author} {\bibfnamefont {Jian}\ \bibnamefont
  {Zhou}},\ }\bibfield  {title} {\enquote {\bibinfo {title}
  {Photo-magnetization in two-dimensional sliding ferroelectrics},}\ }\href
  {\doibase 10.1038/s41699-022-00297-6} {\bibfield  {journal} {\bibinfo
  {journal} {npj 2D Materials and Applications}\ }\textbf {\bibinfo {volume}
  {6}},\ \bibinfo {pages} {15} (\bibinfo {year} {2022})}\BibitemShut {NoStop}%
\bibitem [{\citenamefont {Mendoza}\ \emph {et~al.}(2024)\citenamefont
  {Mendoza}, \citenamefont {Arzate-Plata}, \citenamefont {Tancogne-Dejean},\
  and\ \citenamefont {Fregoso}}]{fregoso_photo24}%
  \BibitemOpen
  \bibfield  {author} {\bibinfo {author} {\bibfnamefont {Bernardo~S.}\
  \bibnamefont {Mendoza}}, \bibinfo {author} {\bibfnamefont {Norberto}\
  \bibnamefont {Arzate-Plata}}, \bibinfo {author} {\bibfnamefont {Nicolas}\
  \bibnamefont {Tancogne-Dejean}}, \ and\ \bibinfo {author} {\bibfnamefont
  {Benjamin~M.}\ \bibnamefont {Fregoso}},\ }\href
  {https://arxiv.org/abs/2406.14748} {\enquote {\bibinfo {title} {Nonlinear
  photomagnetization in insulators},}\ } (\bibinfo {year} {2024}),\ \Eprint
  {http://arxiv.org/abs/2406.14748} {arXiv:2406.14748 [cond-mat.mtrl-sci]}
  \BibitemShut {NoStop}%
\bibitem [{\citenamefont {Xu}\ \emph {et~al.}(2021)\citenamefont {Xu},
  \citenamefont {Zhou}, \citenamefont {Wang},\ and\ \citenamefont
  {Li}}]{Li_prb21}%
  \BibitemOpen
  \bibfield  {author} {\bibinfo {author} {\bibfnamefont {Haowei}\ \bibnamefont
  {Xu}}, \bibinfo {author} {\bibfnamefont {Jian}\ \bibnamefont {Zhou}},
  \bibinfo {author} {\bibfnamefont {Hua}\ \bibnamefont {Wang}}, \ and\ \bibinfo
  {author} {\bibfnamefont {Ju}~\bibnamefont {Li}},\ }\bibfield  {title}
  {\enquote {\bibinfo {title} {Light-induced static magnetization: Nonlinear
  edelstein effect},}\ }\href {\doibase 10.1103/PhysRevB.103.205417} {\bibfield
   {journal} {\bibinfo  {journal} {Physical Review B}\ }\textbf {\bibinfo
  {volume} {103}},\ \bibinfo {pages} {205417} (\bibinfo {year}
  {2021})}\BibitemShut {NoStop}%
\bibitem [{\citenamefont {Baek}\ \emph {et~al.}(2024)\citenamefont {Baek},
  \citenamefont {Han}, \citenamefont {Cheon},\ and\ \citenamefont
  {Lee}}]{HWLee_npj24}%
  \BibitemOpen
  \bibfield  {author} {\bibinfo {author} {\bibfnamefont {Insu}\ \bibnamefont
  {Baek}}, \bibinfo {author} {\bibfnamefont {Seungyun}\ \bibnamefont {Han}},
  \bibinfo {author} {\bibfnamefont {Suik}\ \bibnamefont {Cheon}}, \ and\
  \bibinfo {author} {\bibfnamefont {Hyun-Woo}\ \bibnamefont {Lee}},\ }\bibfield
   {title} {\enquote {\bibinfo {title} {Nonlinear orbital and spin edelstein
  effect in centrosymmetric metals},}\ }\href
  {http://dx.doi.org/10.1038/s44306-024-00041-4} {\bibfield  {journal}
  {\bibinfo  {journal} {npj Spintronics}\ }\textbf {\bibinfo {volume} {2}}
  (\bibinfo {year} {2024})}\BibitemShut {NoStop}%
\bibitem [{\citenamefont {Fang}\ \emph {et~al.}(2024)\citenamefont {Fang},
  \citenamefont {Wu}, \citenamefont {Zhang}, \citenamefont {Li},\ and\
  \citenamefont {Zhou}}]{Fang_24_perpective}%
  \BibitemOpen
  \bibfield  {author} {\bibinfo {author} {\bibfnamefont {Ning}\ \bibnamefont
  {Fang}}, \bibinfo {author} {\bibfnamefont {Changqing}\ \bibnamefont {Wu}},
  \bibinfo {author} {\bibfnamefont {Yuzhe}\ \bibnamefont {Zhang}}, \bibinfo
  {author} {\bibfnamefont {Zhongyu}\ \bibnamefont {Li}}, \ and\ \bibinfo
  {author} {\bibfnamefont {Ziyao}\ \bibnamefont {Zhou}},\ }\bibfield  {title}
  {\enquote {\bibinfo {title} {Perspectives: Light control of magnetism and
  device development},}\ }\href {\doibase 10.1021/acsnano.3c13002} {\bibfield
  {journal} {\bibinfo  {journal} {ACS Nano}\ }\textbf {\bibinfo {volume}
  {18}},\ \bibinfo {pages} {8600–8625} (\bibinfo {year} {2024})}\BibitemShut
  {NoStop}%
\bibitem [{\citenamefont {Wu}\ \emph {et~al.}(2023)\citenamefont {Wu},
  \citenamefont {Zhang}, \citenamefont {Wang},\ and\ \citenamefont
  {Meng}}]{Wu_23ultra}%
  \BibitemOpen
  \bibfield  {author} {\bibinfo {author} {\bibfnamefont {Na}~\bibnamefont
  {Wu}}, \bibinfo {author} {\bibfnamefont {Shengjie}\ \bibnamefont {Zhang}},
  \bibinfo {author} {\bibfnamefont {Yaxian}\ \bibnamefont {Wang}}, \ and\
  \bibinfo {author} {\bibfnamefont {Sheng}\ \bibnamefont {Meng}},\ }\bibfield
  {title} {\enquote {\bibinfo {title} {Ultrafast all-optical quantum control of
  magnetization dynamics},}\ }\href {\doibase 10.1016/j.progsurf.2023.100709}
  {\bibfield  {journal} {\bibinfo  {journal} {Progress in Surface Science}\
  }\textbf {\bibinfo {volume} {98}},\ \bibinfo {pages} {100709} (\bibinfo
  {year} {2023})}\BibitemShut {NoStop}%
\bibitem [{\citenamefont {Hamamera}\ \emph {et~al.}(2022)\citenamefont
  {Hamamera}, \citenamefont {Guimarães}, \citenamefont {dos Santos~Dias},\
  and\ \citenamefont {Lounis}}]{Hamamera_cp22}%
  \BibitemOpen
  \bibfield  {author} {\bibinfo {author} {\bibfnamefont {Hanan}\ \bibnamefont
  {Hamamera}}, \bibinfo {author} {\bibfnamefont {Filipe Souza~Mendes}\
  \bibnamefont {Guimarães}}, \bibinfo {author} {\bibfnamefont {Manuel}\
  \bibnamefont {dos Santos~Dias}}, \ and\ \bibinfo {author} {\bibfnamefont
  {Samir}\ \bibnamefont {Lounis}},\ }\bibfield  {title} {\enquote {\bibinfo
  {title} {Polarisation-dependent single-pulse ultrafast optical switching of
  an elementary ferromagnet},}\ }\href
  {http://dx.doi.org/10.1038/s42005-021-00798-8} {\bibfield  {journal}
  {\bibinfo  {journal} {Communications Physics}\ }\textbf {\bibinfo {volume}
  {5}} (\bibinfo {year} {2022})}\BibitemShut {NoStop}%
\bibitem [{\citenamefont {Wang}\ and\ \citenamefont
  {Liu}(2020)}]{Wang_NConv20}%
  \BibitemOpen
  \bibfield  {author} {\bibinfo {author} {\bibfnamefont {Chuangtang}\
  \bibnamefont {Wang}}\ and\ \bibinfo {author} {\bibfnamefont {Yongmin}\
  \bibnamefont {Liu}},\ }\bibfield  {title} {\enquote {\bibinfo {title}
  {Ultrafast optical manipulation of magnetic order in ferromagnetic
  materials},}\ }\href {http://dx.doi.org/10.1186/s40580-020-00246-3}
  {\bibfield  {journal} {\bibinfo  {journal} {Nano Convergence}\ }\textbf
  {\bibinfo {volume} {7}} (\bibinfo {year} {2020})}\BibitemShut {NoStop}%
\bibitem [{\citenamefont {Gambardella}\ and\ \citenamefont
  {Miron}(2011)}]{Gambardella2011}%
  \BibitemOpen
  \bibfield  {author} {\bibinfo {author} {\bibfnamefont {Pietro}\ \bibnamefont
  {Gambardella}}\ and\ \bibinfo {author} {\bibfnamefont {Ioan~Mihai}\
  \bibnamefont {Miron}},\ }\bibfield  {title} {\enquote {\bibinfo {title}
  {Current-induced spin–orbit torques},}\ }\href {\doibase
  10.1098/rsta.2010.0336} {\bibfield  {journal} {\bibinfo  {journal}
  {Philosophical Transactions of the Royal Society A: Mathematical, Physical
  and Engineering Sciences}\ }\textbf {\bibinfo {volume} {369}},\ \bibinfo
  {pages} {3175–3197} (\bibinfo {year} {2011})}\BibitemShut {NoStop}%
\bibitem [{\citenamefont {\ifmmode~\check{Z}\else \v{Z}\fi{}elezn\'y}\ \emph
  {et~al.}(2014{\natexlab{b}})\citenamefont {\ifmmode~\check{Z}\else
  \v{Z}\fi{}elezn\'y}, \citenamefont {Gao}, \citenamefont {V\'yborn\'y},
  \citenamefont {Zemen}, \citenamefont {Ma\ifmmode~\check{s}\else
  \v{s}\fi{}ek}, \citenamefont {Manchon}, \citenamefont {Wunderlich},
  \citenamefont {Sinova},\ and\ \citenamefont {Jungwirth}}]{Jungwirth_prl14}%
  \BibitemOpen
  \bibfield  {author} {\bibinfo {author} {\bibfnamefont {J.}~\bibnamefont
  {\ifmmode~\check{Z}\else \v{Z}\fi{}elezn\'y}}, \bibinfo {author}
  {\bibfnamefont {H.}~\bibnamefont {Gao}}, \bibinfo {author} {\bibfnamefont
  {K.}~\bibnamefont {V\'yborn\'y}}, \bibinfo {author} {\bibfnamefont
  {J.}~\bibnamefont {Zemen}}, \bibinfo {author} {\bibfnamefont
  {J.}~\bibnamefont {Ma\ifmmode~\check{s}\else \v{s}\fi{}ek}}, \bibinfo
  {author} {\bibfnamefont {Aur\'elien}\ \bibnamefont {Manchon}}, \bibinfo
  {author} {\bibfnamefont {J.}~\bibnamefont {Wunderlich}}, \bibinfo {author}
  {\bibfnamefont {Jairo}\ \bibnamefont {Sinova}}, \ and\ \bibinfo {author}
  {\bibfnamefont {T.}~\bibnamefont {Jungwirth}},\ }\bibfield  {title} {\enquote
  {\bibinfo {title} {Relativistic n\'eel-order fields induced by electrical
  current in antiferromagnets},}\ }\href {\doibase
  10.1103/PhysRevLett.113.157201} {\bibfield  {journal} {\bibinfo  {journal}
  {Physical Review Letters}\ }\textbf {\bibinfo {volume} {113}},\ \bibinfo
  {pages} {157201} (\bibinfo {year} {2014}{\natexlab{b}})}\BibitemShut
  {NoStop}%
\bibitem [{\citenamefont {\ifmmode~\check{Z}\else \v{Z}\fi{}elezn\'y}\ \emph
  {et~al.}(2017{\natexlab{b}})\citenamefont {\ifmmode~\check{Z}\else
  \v{Z}\fi{}elezn\'y}, \citenamefont {Gao}, \citenamefont {Manchon},
  \citenamefont {Freimuth}, \citenamefont {Mokrousov}, \citenamefont {Zemen},
  \citenamefont {Ma\ifmmode~\check{s}\else \v{s}\fi{}ek}, \citenamefont
  {Sinova},\ and\ \citenamefont {Jungwirth}}]{Jungwirth_prb17}%
  \BibitemOpen
  \bibfield  {author} {\bibinfo {author} {\bibfnamefont {J.}~\bibnamefont
  {\ifmmode~\check{Z}\else \v{Z}\fi{}elezn\'y}}, \bibinfo {author}
  {\bibfnamefont {H.}~\bibnamefont {Gao}}, \bibinfo {author} {\bibfnamefont
  {Aur\'elien}\ \bibnamefont {Manchon}}, \bibinfo {author} {\bibfnamefont
  {Frank}\ \bibnamefont {Freimuth}}, \bibinfo {author} {\bibfnamefont {Yuriy}\
  \bibnamefont {Mokrousov}}, \bibinfo {author} {\bibfnamefont {J.}~\bibnamefont
  {Zemen}}, \bibinfo {author} {\bibfnamefont {J.}~\bibnamefont
  {Ma\ifmmode~\check{s}\else \v{s}\fi{}ek}}, \bibinfo {author} {\bibfnamefont
  {Jairo}\ \bibnamefont {Sinova}}, \ and\ \bibinfo {author} {\bibfnamefont
  {T.}~\bibnamefont {Jungwirth}},\ }\bibfield  {title} {\enquote {\bibinfo
  {title} {Spin-orbit torques in locally and globally noncentrosymmetric
  crystals: Antiferromagnets and ferromagnets},}\ }\href {\doibase
  10.1103/PhysRevB.95.014403} {\bibfield  {journal} {\bibinfo  {journal}
  {Physical Review B}\ }\textbf {\bibinfo {volume} {95}},\ \bibinfo {pages}
  {014403} (\bibinfo {year} {2017}{\natexlab{b}})}\BibitemShut {NoStop}%
\bibitem [{\citenamefont {Feng}\ \emph
  {et~al.}(2024{\natexlab{b}})\citenamefont {Feng}, \citenamefont {Wu},
  \citenamefont {Wang}, \citenamefont {Gao}, \citenamefont {Ang}, \citenamefont
  {Zhao}, \citenamefont {Xiao},\ and\ \citenamefont {Yang}}]{Yang__24quantum}%
  \BibitemOpen
  \bibfield  {author} {\bibinfo {author} {\bibfnamefont {Xukun}\ \bibnamefont
  {Feng}}, \bibinfo {author} {\bibfnamefont {Weikang}\ \bibnamefont {Wu}},
  \bibinfo {author} {\bibfnamefont {Hui}\ \bibnamefont {Wang}}, \bibinfo
  {author} {\bibfnamefont {Weibo}\ \bibnamefont {Gao}}, \bibinfo {author}
  {\bibfnamefont {Lay~Kee}\ \bibnamefont {Ang}}, \bibinfo {author}
  {\bibfnamefont {Y.~X.}\ \bibnamefont {Zhao}}, \bibinfo {author}
  {\bibfnamefont {Cong}\ \bibnamefont {Xiao}}, \ and\ \bibinfo {author}
  {\bibfnamefont {Shengyuan~A.}\ \bibnamefont {Yang}},\ }\href
  {https://arxiv.org/abs/2402.00532} {\enquote {\bibinfo {title} {Quantum
  metric nonlinear spin-orbit torque enhanced by topological bands},}\ }
  (\bibinfo {year} {2024}{\natexlab{b}}),\ \Eprint
  {http://arxiv.org/abs/2402.00532} {arXiv:2402.00532 [cond-mat.mes-hall]}
  \BibitemShut {NoStop}%
\bibitem [{\citenamefont {Zhou}\ \emph {et~al.}(2022)\citenamefont {Zhou},
  \citenamefont {Duan}, \citenamefont {Wu}, \citenamefont {Deng}, \citenamefont
  {Wang}, \citenamefont {Culcer},\ and\ \citenamefont {Wang}}]{Wang_prb22}%
  \BibitemOpen
  \bibfield  {author} {\bibinfo {author} {\bibfnamefont {Yong-Long}\
  \bibnamefont {Zhou}}, \bibinfo {author} {\bibfnamefont {Hou-Jian}\
  \bibnamefont {Duan}}, \bibinfo {author} {\bibfnamefont {Yong-jia}\
  \bibnamefont {Wu}}, \bibinfo {author} {\bibfnamefont {Ming-Xun}\ \bibnamefont
  {Deng}}, \bibinfo {author} {\bibfnamefont {Lan}\ \bibnamefont {Wang}},
  \bibinfo {author} {\bibfnamefont {Dimitrie}\ \bibnamefont {Culcer}}, \ and\
  \bibinfo {author} {\bibfnamefont {Rui-Qiang}\ \bibnamefont {Wang}},\
  }\bibfield  {title} {\enquote {\bibinfo {title} {Nonlinear antidamping
  spin-orbit torque originating from intraband transport on the warped surface
  of a topological insulator},}\ }\href {\doibase 10.1103/PhysRevB.105.075415}
  {\bibfield  {journal} {\bibinfo  {journal} {Physical Review B}\ }\textbf
  {\bibinfo {volume} {105}},\ \bibinfo {pages} {075415} (\bibinfo {year}
  {2022})}\BibitemShut {NoStop}%
\bibitem [{\citenamefont {Cheng}\ \emph {et~al.}(2016)\citenamefont {Cheng},
  \citenamefont {Xiao},\ and\ \citenamefont {Brataas}}]{Cheng_prl15}%
  \BibitemOpen
  \bibfield  {author} {\bibinfo {author} {\bibfnamefont {Ran}\ \bibnamefont
  {Cheng}}, \bibinfo {author} {\bibfnamefont {Di}~\bibnamefont {Xiao}}, \ and\
  \bibinfo {author} {\bibfnamefont {Arne}\ \bibnamefont {Brataas}},\ }\bibfield
   {title} {\enquote {\bibinfo {title} {Terahertz antiferromagnetic spin hall
  nano-oscillator},}\ }\href {\doibase 10.1103/PhysRevLett.116.207603}
  {\bibfield  {journal} {\bibinfo  {journal} {Physical Review Letters}\
  }\textbf {\bibinfo {volume} {116}},\ \bibinfo {pages} {207603} (\bibinfo
  {year} {2016})}\BibitemShut {NoStop}%
\bibitem [{\citenamefont {Khymyn}\ \emph {et~al.}(2017)\citenamefont {Khymyn},
  \citenamefont {Lisenkov}, \citenamefont {Tiberkevich}, \citenamefont
  {Ivanov},\ and\ \citenamefont {Slavin}}]{Khymyn_nature17}%
  \BibitemOpen
  \bibfield  {author} {\bibinfo {author} {\bibfnamefont {Roman}\ \bibnamefont
  {Khymyn}}, \bibinfo {author} {\bibfnamefont {Ivan}\ \bibnamefont {Lisenkov}},
  \bibinfo {author} {\bibfnamefont {Vasyl}\ \bibnamefont {Tiberkevich}},
  \bibinfo {author} {\bibfnamefont {Boris~A.}\ \bibnamefont {Ivanov}}, \ and\
  \bibinfo {author} {\bibfnamefont {Andrei}\ \bibnamefont {Slavin}},\
  }\bibfield  {title} {\enquote {\bibinfo {title} {Antiferromagnetic
  thz-frequency josephson-like oscillator driven by spin current},}\ }\href
  {\doibase 10.1038/srep43705} {\bibfield  {journal} {\bibinfo  {journal}
  {Scientific Reports}\ }\textbf {\bibinfo {volume} {7}},\ \bibinfo {pages}
  {43705} (\bibinfo {year} {2017})}\BibitemShut {NoStop}%
\bibitem [{\citenamefont {Sekine}\ \emph {et~al.}(2017)\citenamefont {Sekine},
  \citenamefont {Culcer},\ and\ \citenamefont {MacDonald}}]{Sekine_prb17}%
  \BibitemOpen
  \bibfield  {author} {\bibinfo {author} {\bibfnamefont {Akihiko}\ \bibnamefont
  {Sekine}}, \bibinfo {author} {\bibfnamefont {Dimitrie}\ \bibnamefont
  {Culcer}}, \ and\ \bibinfo {author} {\bibfnamefont {Allan~H.}\ \bibnamefont
  {MacDonald}},\ }\bibfield  {title} {\enquote {\bibinfo {title} {Quantum
  kinetic theory of the chiral anomaly},}\ }\href {\doibase
  10.1103/PhysRevB.96.235134} {\bibfield  {journal} {\bibinfo  {journal}
  {Physical Review B}\ }\textbf {\bibinfo {volume} {96}},\ \bibinfo {pages}
  {235134} (\bibinfo {year} {2017})}\BibitemShut {NoStop}%
\bibitem [{Note1()}]{Note1}%
  \BibitemOpen
  \bibinfo {note} {In principle, the magnetic field ${\protect \bm {B}}(t)$ of
  the electromagnetic field should also be considered in the light-matter
  coupling term. However, the magnetic coupling is very weak compared to the
  electric field of the light as $B \sim E/v_0$, with $v_0$ being the velocity
  of light in the material in discussion. Consequently, the spin-Zeeman
  coupling term in the light-matter interaction Hamiltonian is very weak
  compared to the electric field-induced term.}\BibitemShut {Stop}%
\bibitem [{\citenamefont {Mandal}\ \emph {et~al.}(2024)\citenamefont {Mandal},
  \citenamefont {Sarkar}, \citenamefont {Das},\ and\ \citenamefont
  {Agarwal}}]{mandal_24quantum}%
  \BibitemOpen
  \bibfield  {author} {\bibinfo {author} {\bibfnamefont {Debottam}\
  \bibnamefont {Mandal}}, \bibinfo {author} {\bibfnamefont {Sanjay}\
  \bibnamefont {Sarkar}}, \bibinfo {author} {\bibfnamefont {Kamal}\
  \bibnamefont {Das}}, \ and\ \bibinfo {author} {\bibfnamefont {Amit}\
  \bibnamefont {Agarwal}},\ }\href {https://arxiv.org/abs/2310.19092} {\enquote
  {\bibinfo {title} {Quantum geometry induced third order nonlinear transport
  responses},}\ } (\bibinfo {year} {2024}),\ \Eprint
  {http://arxiv.org/abs/2310.19092} {arXiv:2310.19092 [cond-mat.mes-hall]}
  \BibitemShut {NoStop}%
\bibitem [{\citenamefont {Varshney}\ \emph {et~al.}(2023)\citenamefont
  {Varshney}, \citenamefont {Das}, \citenamefont {Bhalla},\ and\ \citenamefont
  {Agarwal}}]{Varshney_2023}%
  \BibitemOpen
  \bibfield  {author} {\bibinfo {author} {\bibfnamefont {Harsh}\ \bibnamefont
  {Varshney}}, \bibinfo {author} {\bibfnamefont {Kamal}\ \bibnamefont {Das}},
  \bibinfo {author} {\bibfnamefont {Pankaj}\ \bibnamefont {Bhalla}}, \ and\
  \bibinfo {author} {\bibfnamefont {Amit}\ \bibnamefont {Agarwal}},\ }\bibfield
   {title} {\enquote {\bibinfo {title} {Quantum kinetic theory of nonlinear
  thermal current},}\ }\href {\doibase 10.1103/PhysRevB.107.235419} {\bibfield
  {journal} {\bibinfo  {journal} {Physical Review B}\ }\textbf {\bibinfo
  {volume} {107}},\ \bibinfo {pages} {235419} (\bibinfo {year}
  {2023})}\BibitemShut {NoStop}%
\bibitem [{\citenamefont {Das}\ \emph {et~al.}(2023)\citenamefont {Das},
  \citenamefont {Lahiri}, \citenamefont {Atencia}, \citenamefont {Culcer},\
  and\ \citenamefont {Agarwal}}]{Das_Lahiri_2023}%
  \BibitemOpen
  \bibfield  {author} {\bibinfo {author} {\bibfnamefont {Kamal}\ \bibnamefont
  {Das}}, \bibinfo {author} {\bibfnamefont {Shibalik}\ \bibnamefont {Lahiri}},
  \bibinfo {author} {\bibfnamefont {Rhonald~Burgos}\ \bibnamefont {Atencia}},
  \bibinfo {author} {\bibfnamefont {Dimitrie}\ \bibnamefont {Culcer}}, \ and\
  \bibinfo {author} {\bibfnamefont {Amit}\ \bibnamefont {Agarwal}},\ }\bibfield
   {title} {\enquote {\bibinfo {title} {Intrinsic nonlinear conductivities
  induced by the quantum metric},}\ }\href {\doibase
  10.1103/PhysRevB.108.L201405} {\bibfield  {journal} {\bibinfo  {journal}
  {Physical Review B}\ }\textbf {\bibinfo {volume} {108}},\ \bibinfo {pages}
  {L201405} (\bibinfo {year} {2023})}\BibitemShut {NoStop}%
\bibitem [{\citenamefont {Kumar}\ \emph {et~al.}(2024)\citenamefont {Kumar},
  \citenamefont {Sarkar},\ and\ \citenamefont {Agarwal}}]{Maneesh_24band}%
  \BibitemOpen
  \bibfield  {author} {\bibinfo {author} {\bibfnamefont {M.~Maneesh}\
  \bibnamefont {Kumar}}, \bibinfo {author} {\bibfnamefont {Sanjay}\
  \bibnamefont {Sarkar}}, \ and\ \bibinfo {author} {\bibfnamefont {Amit}\
  \bibnamefont {Agarwal}},\ }\bibfield  {title} {\enquote {\bibinfo {title}
  {Band geometry induced electro-optic effect and polarization rotation},}\
  }\href {\doibase 10.1103/PhysRevB.110.125401} {\bibfield  {journal} {\bibinfo
   {journal} {Physical Review B}\ }\textbf {\bibinfo {volume} {110}},\ \bibinfo
  {pages} {125401} (\bibinfo {year} {2024})}\BibitemShut {NoStop}%
\bibitem [{Note2()}]{Note2}%
  \BibitemOpen
  \bibinfo {note} {The Supplemental Material discuss: S1) Detailed calculation
  of light-induced spin magnetization; S2) Gauge invariance of spin
  susceptibilities, S3) Spin susceptibilities under different light
  polarizations; S4) Angular dependence of spin magnetization under linearly
  polarized light; S5) Detailed symmetry analysis; S6) Semimetallic
  antiferromagnet: quasi-2D model of {\ch {CuMnAs}}; S7) Applications of
  nonlinear spin magnetization.}\BibitemShut {Stop}%
\bibitem [{\citenamefont {Sodemann}\ and\ \citenamefont
  {Fu}(2015)}]{Sodemann_prl15}%
  \BibitemOpen
  \bibfield  {author} {\bibinfo {author} {\bibfnamefont {Inti}\ \bibnamefont
  {Sodemann}}\ and\ \bibinfo {author} {\bibfnamefont {Liang}\ \bibnamefont
  {Fu}},\ }\bibfield  {title} {\enquote {\bibinfo {title} {Quantum nonlinear
  hall effect induced by berry curvature dipole in time-reversal invariant
  materials},}\ }\href {\doibase 10.1103/PhysRevLett.115.216806} {\bibfield
  {journal} {\bibinfo  {journal} {Physical Review Letters}\ }\textbf {\bibinfo
  {volume} {115}},\ \bibinfo {pages} {216806} (\bibinfo {year}
  {2015})}\BibitemShut {NoStop}%
\bibitem [{Note3()}]{Note3}%
  \BibitemOpen
  \bibinfo {note} {This is because $\protect \tilde {\alpha }_{a;bc}^{\protect
  \rm SSh}$ contains shift-vector ($A^{bc}_{mp}$) in the form of $D_{mp}^b
  \protect \mathcal {R}_{mp}^c=-i{\protect \cal R}_{mp}^c A^{bc}_{mp}$~\cite
  {Sipe_prb00,Bhalla_2022}, which signifies the position shift of real-space
  wavefunction of Bloch electrons as shown in Fig.~\ref
  {Fig_1}(b).}\BibitemShut {Stop}%
\bibitem [{\citenamefont {Sipe}\ and\ \citenamefont
  {Shkrebtii}(2000)}]{Sipe_prb00}%
  \BibitemOpen
  \bibfield  {author} {\bibinfo {author} {\bibfnamefont {J.~E.}\ \bibnamefont
  {Sipe}}\ and\ \bibinfo {author} {\bibfnamefont {A.~I.}\ \bibnamefont
  {Shkrebtii}},\ }\bibfield  {title} {\enquote {\bibinfo {title} {Second-order
  optical response in semiconductors},}\ }\href {\doibase
  10.1103/PhysRevB.61.5337} {\bibfield  {journal} {\bibinfo  {journal}
  {Physical Review B}\ }\textbf {\bibinfo {volume} {61}},\ \bibinfo {pages}
  {5337--5352} (\bibinfo {year} {2000})}\BibitemShut {NoStop}%
\bibitem [{\citenamefont {Aversa}\ and\ \citenamefont
  {Sipe}(1995)}]{Aversa_1995}%
  \BibitemOpen
  \bibfield  {author} {\bibinfo {author} {\bibfnamefont {Claudio}\ \bibnamefont
  {Aversa}}\ and\ \bibinfo {author} {\bibfnamefont {J.~E.}\ \bibnamefont
  {Sipe}},\ }\bibfield  {title} {\enquote {\bibinfo {title} {Nonlinear optical
  susceptibilities of semiconductors: Results with a length-gauge analysis},}\
  }\href {\doibase 10.1103/PhysRevB.52.14636} {\bibfield  {journal} {\bibinfo
  {journal} {Physical Review B}\ }\textbf {\bibinfo {volume} {52}},\ \bibinfo
  {pages} {14636--14645} (\bibinfo {year} {1995})}\BibitemShut {NoStop}%
\bibitem [{\citenamefont {Bhalla}\ \emph {et~al.}(2022)\citenamefont {Bhalla},
  \citenamefont {Das}, \citenamefont {Culcer},\ and\ \citenamefont
  {Agarwal}}]{Bhalla_2022}%
  \BibitemOpen
  \bibfield  {author} {\bibinfo {author} {\bibfnamefont {Pankaj}\ \bibnamefont
  {Bhalla}}, \bibinfo {author} {\bibfnamefont {Kamal}\ \bibnamefont {Das}},
  \bibinfo {author} {\bibfnamefont {Dimitrie}\ \bibnamefont {Culcer}}, \ and\
  \bibinfo {author} {\bibfnamefont {Amit}\ \bibnamefont {Agarwal}},\ }\bibfield
   {title} {\enquote {\bibinfo {title} {Resonant second-harmonic generation as
  a probe of quantum geometry},}\ }\href
  {https://doi.org/10.1103%2Fphysrevlett.129.227401} {\bibfield  {journal}
  {\bibinfo  {journal} {Physical Review Letters}\ }\textbf {\bibinfo {volume}
  {129}} (\bibinfo {year} {2022})}\BibitemShut {NoStop}%
\bibitem [{Note4()}]{Note4}%
  \BibitemOpen
  \bibinfo {note} {We find that for rectification response of $\alpha _{a;bc}$,
  only Drude susceptibility is completely $\protect \cal T$-odd, all other
  susceptibilities have both $\protect \cal T$-even and $\protect \cal T$-odd
  components. In contrast, for second harmonic response, only the Drude and
  spin injection is completely $\protect \cal T$-odd. All other components have
  both $\protect \cal T$ odd and $\protect \cal T$-even
  counterparts.}\BibitemShut {Stop}%
\bibitem [{\citenamefont {\ifmmode~\check{S}\else \v{S}\fi{}mejkal}\ \emph
  {et~al.}(2017)\citenamefont {\ifmmode~\check{S}\else \v{S}\fi{}mejkal},
  \citenamefont {\ifmmode~\check{Z}\else \v{Z}\fi{}elezn\'y}, \citenamefont
  {Sinova},\ and\ \citenamefont {Jungwirth}}]{smejkal_prl17}%
  \BibitemOpen
  \bibfield  {author} {\bibinfo {author} {\bibfnamefont {L.}~\bibnamefont
  {\ifmmode~\check{S}\else \v{S}\fi{}mejkal}}, \bibinfo {author} {\bibfnamefont
  {J.}~\bibnamefont {\ifmmode~\check{Z}\else \v{Z}\fi{}elezn\'y}}, \bibinfo
  {author} {\bibfnamefont {J.}~\bibnamefont {Sinova}}, \ and\ \bibinfo {author}
  {\bibfnamefont {T.}~\bibnamefont {Jungwirth}},\ }\bibfield  {title} {\enquote
  {\bibinfo {title} {Electric control of dirac quasiparticles by spin-orbit
  torque in an antiferromagnet},}\ }\href {\doibase
  10.1103/PhysRevLett.118.106402} {\bibfield  {journal} {\bibinfo  {journal}
  {Physical Review Letters}\ }\textbf {\bibinfo {volume} {118}},\ \bibinfo
  {pages} {106402} (\bibinfo {year} {2017})}\BibitemShut {NoStop}%
\bibitem [{\citenamefont {Watanabe}\ and\ \citenamefont
  {Yanase}(2021)}]{Watanabe_prx21}%
  \BibitemOpen
  \bibfield  {author} {\bibinfo {author} {\bibfnamefont {Hikaru}\ \bibnamefont
  {Watanabe}}\ and\ \bibinfo {author} {\bibfnamefont {Youichi}\ \bibnamefont
  {Yanase}},\ }\bibfield  {title} {\enquote {\bibinfo {title} {Chiral
  photocurrent in parity-violating magnet and enhanced response in topological
  antiferromagnet},}\ }\href {\doibase 10.1103/PhysRevX.11.011001} {\bibfield
  {journal} {\bibinfo  {journal} {Physical Review X}\ }\textbf {\bibinfo
  {volume} {11}},\ \bibinfo {pages} {011001} (\bibinfo {year}
  {2021})}\BibitemShut {NoStop}%
\bibitem [{Note5()}]{Note5}%
  \BibitemOpen
  \bibinfo {note} {Here, we have converted the $\delta S^{(2)}$ in conventional
  3D units by dividing the result by the thickness $\sim 6.38$ \r A~\cite
  {Wadley_sr2015} of the CuMnAs unit cell.}\BibitemShut {Stop}%
\bibitem [{\citenamefont {Fang}\ \emph {et~al.}(2011)\citenamefont {Fang},
  \citenamefont {Kurebayashi}, \citenamefont {Wunderlich}, \citenamefont
  {Výborný}, \citenamefont {Z\^arbo}, \citenamefont {Campion}, \citenamefont
  {Casiraghi}, \citenamefont {Gallagher}, \citenamefont {Jungwirth},\ and\
  \citenamefont {Ferguson}}]{Fang_NN11}%
  \BibitemOpen
  \bibfield  {author} {\bibinfo {author} {\bibfnamefont {D.}~\bibnamefont
  {Fang}}, \bibinfo {author} {\bibfnamefont {H.}~\bibnamefont {Kurebayashi}},
  \bibinfo {author} {\bibfnamefont {J.}~\bibnamefont {Wunderlich}}, \bibinfo
  {author} {\bibfnamefont {K.}~\bibnamefont {Výborný}}, \bibinfo {author}
  {\bibfnamefont {L.~P.}\ \bibnamefont {Z\^arbo}}, \bibinfo {author}
  {\bibfnamefont {R.~P.}\ \bibnamefont {Campion}}, \bibinfo {author}
  {\bibfnamefont {A.}~\bibnamefont {Casiraghi}}, \bibinfo {author}
  {\bibfnamefont {B.~L.}\ \bibnamefont {Gallagher}}, \bibinfo {author}
  {\bibfnamefont {T.}~\bibnamefont {Jungwirth}}, \ and\ \bibinfo {author}
  {\bibfnamefont {A.~J.}\ \bibnamefont {Ferguson}},\ }\bibfield  {title}
  {\enquote {\bibinfo {title} {Spin–orbit-driven ferromagnetic resonance},}\
  }\href {\doibase 10.1038/nnano.2011.68} {\bibfield  {journal} {\bibinfo
  {journal} {Nature Nanotechnology}\ }\textbf {\bibinfo {volume} {6}},\
  \bibinfo {pages} {413–417} (\bibinfo {year} {2011})}\BibitemShut {NoStop}%
\bibitem [{\citenamefont {Kurebayashi}\ \emph {et~al.}(2014)\citenamefont
  {Kurebayashi}, \citenamefont {Sinova}, \citenamefont {Fang}, \citenamefont
  {Irvine}, \citenamefont {Skinner}, \citenamefont {Wunderlich}, \citenamefont
  {Novák}, \citenamefont {Campion}, \citenamefont {Gallagher}, \citenamefont
  {Vehstedt}, \citenamefont {Z\^arbo}, \citenamefont {Výborný}, \citenamefont
  {Ferguson},\ and\ \citenamefont {Jungwirth}}]{Kurebayashi_NN14}%
  \BibitemOpen
  \bibfield  {author} {\bibinfo {author} {\bibfnamefont {H.}~\bibnamefont
  {Kurebayashi}}, \bibinfo {author} {\bibfnamefont {Jairo}\ \bibnamefont
  {Sinova}}, \bibinfo {author} {\bibfnamefont {D.}~\bibnamefont {Fang}},
  \bibinfo {author} {\bibfnamefont {A.~C.}\ \bibnamefont {Irvine}}, \bibinfo
  {author} {\bibfnamefont {T.~D.}\ \bibnamefont {Skinner}}, \bibinfo {author}
  {\bibfnamefont {J.}~\bibnamefont {Wunderlich}}, \bibinfo {author}
  {\bibfnamefont {V.}~\bibnamefont {Novák}}, \bibinfo {author} {\bibfnamefont
  {R.~P.}\ \bibnamefont {Campion}}, \bibinfo {author} {\bibfnamefont {B.~L.}\
  \bibnamefont {Gallagher}}, \bibinfo {author} {\bibfnamefont {E.~K.}\
  \bibnamefont {Vehstedt}}, \bibinfo {author} {\bibfnamefont {L.~P.}\
  \bibnamefont {Z\^arbo}}, \bibinfo {author} {\bibfnamefont {K.}~\bibnamefont
  {Výborný}}, \bibinfo {author} {\bibfnamefont {A.~J.}\ \bibnamefont
  {Ferguson}}, \ and\ \bibinfo {author} {\bibfnamefont {T.}~\bibnamefont
  {Jungwirth}},\ }\bibfield  {title} {\enquote {\bibinfo {title} {An
  antidamping spin–orbit torque originating from the berry curvature},}\
  }\href {\doibase 10.1038/nnano.2014.15} {\bibfield  {journal} {\bibinfo
  {journal} {Nature Nanotechnology}\ }\textbf {\bibinfo {volume} {9}},\
  \bibinfo {pages} {211–217} (\bibinfo {year} {2014})}\BibitemShut {NoStop}%
\bibitem [{\citenamefont {Chernyshov}\ \emph {et~al.}(2009)\citenamefont
  {Chernyshov}, \citenamefont {Overby}, \citenamefont {Liu}, \citenamefont
  {Furdyna}, \citenamefont {Lyanda-Geller},\ and\ \citenamefont
  {Rokhinson}}]{Chernyshov_np09}%
  \BibitemOpen
  \bibfield  {author} {\bibinfo {author} {\bibfnamefont {Alexandr}\
  \bibnamefont {Chernyshov}}, \bibinfo {author} {\bibfnamefont {Mason}\
  \bibnamefont {Overby}}, \bibinfo {author} {\bibfnamefont {Xinyu}\
  \bibnamefont {Liu}}, \bibinfo {author} {\bibfnamefont {Jacek~K.}\
  \bibnamefont {Furdyna}}, \bibinfo {author} {\bibfnamefont {Yuli}\
  \bibnamefont {Lyanda-Geller}}, \ and\ \bibinfo {author} {\bibfnamefont
  {Leonid~P.}\ \bibnamefont {Rokhinson}},\ }\bibfield  {title} {\enquote
  {\bibinfo {title} {Evidence for reversible control of magnetization in a
  ferromagnetic material by means of spin--orbit magnetic field},}\ }\href
  {\doibase 10.1038/nphys1362} {\bibfield  {journal} {\bibinfo  {journal}
  {Nature Physics}\ }\textbf {\bibinfo {volume} {5}},\ \bibinfo {pages}
  {656--659} (\bibinfo {year} {2009})}\BibitemShut {NoStop}%
\bibitem [{Note6()}]{Note6}%
  \BibitemOpen
  \bibinfo {note} {We found the LNSM value of tetragonal CuMnAs in the
  semimetallic state to be around $1.12~\mu _B/{\protect \rm nm^2}$ under LPL
  for an electric field amplitude $1$ MV/m. We mention that linear-order spin
  magnetization in ${\protect \rm CuMnAs}$ is typically around $10^{-14}\mu
  _B/{\protect \rm nm}^2$ for the same field value. In a \ch {CrI3} bilayer,
  spin magnetization of approximately $0.04\mu _B/{\protect \rm nm}^2$ can be
  found for the same laser intensity~\cite {Li_prb21}. Similarly, in a single
  septuple layer of \ch {MnBi2Te4}, the spin magnetization is found to be
  $0.7\times 10^{-3}~\mu _B/{\protect \rm nm^2}$ for the same electric field
  value~\cite {Xiao_2023}.}\BibitemShut {Stop}%
\bibitem [{\citenamefont {Kodama}\ \emph {et~al.}(2024)\citenamefont {Kodama},
  \citenamefont {Kikuchi}, \citenamefont {Chiba}, \citenamefont {Okamoto},
  \citenamefont {Ohno},\ and\ \citenamefont {Tomita}}]{Satohi_prb24}%
  \BibitemOpen
  \bibfield  {author} {\bibinfo {author} {\bibfnamefont {Toshiyuki}\
  \bibnamefont {Kodama}}, \bibinfo {author} {\bibfnamefont {Nobuaki}\
  \bibnamefont {Kikuchi}}, \bibinfo {author} {\bibfnamefont {Takahiro}\
  \bibnamefont {Chiba}}, \bibinfo {author} {\bibfnamefont {Satoshi}\
  \bibnamefont {Okamoto}}, \bibinfo {author} {\bibfnamefont {Seigo}\
  \bibnamefont {Ohno}}, \ and\ \bibinfo {author} {\bibfnamefont {Satoshi}\
  \bibnamefont {Tomita}},\ }\bibfield  {title} {\enquote {\bibinfo {title}
  {Direct observation of current-induced nonlinear spin torque in pt-py
  bilayers},}\ }\href {\doibase 10.1103/PhysRevB.109.214419} {\bibfield
  {journal} {\bibinfo  {journal} {Physical Review B}\ }\textbf {\bibinfo
  {volume} {109}},\ \bibinfo {pages} {214419} (\bibinfo {year}
  {2024})}\BibitemShut {NoStop}%
\bibitem [{\citenamefont {Go}\ and\ \citenamefont {Lee}(2020)}]{HWLee_prb20}%
  \BibitemOpen
  \bibfield  {author} {\bibinfo {author} {\bibfnamefont {Dongwook}\
  \bibnamefont {Go}}\ and\ \bibinfo {author} {\bibfnamefont {Hyun-Woo}\
  \bibnamefont {Lee}},\ }\bibfield  {title} {\enquote {\bibinfo {title}
  {Orbital torque: Torque generation by orbital current injection},}\ }\href
  {\doibase 10.1103/PhysRevResearch.2.013177} {\bibfield  {journal} {\bibinfo
  {journal} {Physical Review Research}\ }\textbf {\bibinfo {volume} {2}},\
  \bibinfo {pages} {013177} (\bibinfo {year} {2020})}\BibitemShut {NoStop}%
\bibitem [{\citenamefont {Lee}\ \emph {et~al.}(2021)\citenamefont {Lee},
  \citenamefont {Go}, \citenamefont {Park}, \citenamefont {Jeong},
  \citenamefont {Ko}, \citenamefont {Yun}, \citenamefont {Jo}, \citenamefont
  {Lee}, \citenamefont {Go}, \citenamefont {Oh}, \citenamefont {Kim},
  \citenamefont {Park}, \citenamefont {Min}, \citenamefont {Koo}, \citenamefont
  {Lee}, \citenamefont {Lee},\ and\ \citenamefont {Lee}}]{Lee_NatComm2021}%
  \BibitemOpen
  \bibfield  {author} {\bibinfo {author} {\bibfnamefont {Dongjoon}\
  \bibnamefont {Lee}}, \bibinfo {author} {\bibfnamefont {Dongwook}\
  \bibnamefont {Go}}, \bibinfo {author} {\bibfnamefont {Hyeon-Jong}\
  \bibnamefont {Park}}, \bibinfo {author} {\bibfnamefont {Wonmin}\ \bibnamefont
  {Jeong}}, \bibinfo {author} {\bibfnamefont {Hye-Won}\ \bibnamefont {Ko}},
  \bibinfo {author} {\bibfnamefont {Deokhyun}\ \bibnamefont {Yun}}, \bibinfo
  {author} {\bibfnamefont {Daegeun}\ \bibnamefont {Jo}}, \bibinfo {author}
  {\bibfnamefont {Soogil}\ \bibnamefont {Lee}}, \bibinfo {author}
  {\bibfnamefont {Gyungchoon}\ \bibnamefont {Go}}, \bibinfo {author}
  {\bibfnamefont {Jung~Hyun}\ \bibnamefont {Oh}}, \bibinfo {author}
  {\bibfnamefont {Kab-Jin}\ \bibnamefont {Kim}}, \bibinfo {author}
  {\bibfnamefont {Byong-Guk}\ \bibnamefont {Park}}, \bibinfo {author}
  {\bibfnamefont {Byoung-Chul}\ \bibnamefont {Min}}, \bibinfo {author}
  {\bibfnamefont {Hyun~Cheol}\ \bibnamefont {Koo}}, \bibinfo {author}
  {\bibfnamefont {Hyun-Woo}\ \bibnamefont {Lee}}, \bibinfo {author}
  {\bibfnamefont {OukJae}\ \bibnamefont {Lee}}, \ and\ \bibinfo {author}
  {\bibfnamefont {Kyung-Jin}\ \bibnamefont {Lee}},\ }\bibfield  {title}
  {\enquote {\bibinfo {title} {Orbital torque in magnetic bilayers},}\ }\href
  {\doibase 10.1038/s41467-021-26650-9} {\bibfield  {journal} {\bibinfo
  {journal} {Nature Communications}\ }\textbf {\bibinfo {volume} {12}},\
  \bibinfo {pages} {6710} (\bibinfo {year} {2021})}\BibitemShut {NoStop}%
\bibitem [{\citenamefont {Fukunaga}\ \emph {et~al.}(2023)\citenamefont
  {Fukunaga}, \citenamefont {Haku}, \citenamefont {Hayashi},\ and\
  \citenamefont {Ando}}]{Ando_PRR23}%
  \BibitemOpen
  \bibfield  {author} {\bibinfo {author} {\bibfnamefont {Riko}\ \bibnamefont
  {Fukunaga}}, \bibinfo {author} {\bibfnamefont {Satoshi}\ \bibnamefont
  {Haku}}, \bibinfo {author} {\bibfnamefont {Hiroki}\ \bibnamefont {Hayashi}},
  \ and\ \bibinfo {author} {\bibfnamefont {Kazuya}\ \bibnamefont {Ando}},\
  }\bibfield  {title} {\enquote {\bibinfo {title} {Orbital torque originating
  from orbital hall effect in zr},}\ }\href {\doibase
  10.1103/PhysRevResearch.5.023054} {\bibfield  {journal} {\bibinfo  {journal}
  {Physical Review Research}\ }\textbf {\bibinfo {volume} {5}},\ \bibinfo
  {pages} {023054} (\bibinfo {year} {2023})}\BibitemShut {NoStop}%
\bibitem [{\citenamefont {Wadley}\ \emph {et~al.}(2015)\citenamefont {Wadley},
  \citenamefont {Hills}, \citenamefont {Shahedkhah}, \citenamefont {Edmonds},
  \citenamefont {Campion}, \citenamefont {Nov{\'a}k}, \citenamefont
  {Ouladdiaf}, \citenamefont {Khalyavin}, \citenamefont {Langridge},
  \citenamefont {Saidl}, \citenamefont {Nemec}, \citenamefont {Rushforth},
  \citenamefont {Gallagher}, \citenamefont {Dhesi}, \citenamefont
  {Maccherozzi}, \citenamefont {{\v{Z}}elezn{\'y}},\ and\ \citenamefont
  {Jungwirth}}]{Wadley_sr2015}%
  \BibitemOpen
  \bibfield  {author} {\bibinfo {author} {\bibfnamefont {P.}~\bibnamefont
  {Wadley}}, \bibinfo {author} {\bibfnamefont {V.}~\bibnamefont {Hills}},
  \bibinfo {author} {\bibfnamefont {M.~R.}\ \bibnamefont {Shahedkhah}},
  \bibinfo {author} {\bibfnamefont {K.~W.}\ \bibnamefont {Edmonds}}, \bibinfo
  {author} {\bibfnamefont {R.~P.}\ \bibnamefont {Campion}}, \bibinfo {author}
  {\bibfnamefont {V.}~\bibnamefont {Nov{\'a}k}}, \bibinfo {author}
  {\bibfnamefont {B.}~\bibnamefont {Ouladdiaf}}, \bibinfo {author}
  {\bibfnamefont {D.}~\bibnamefont {Khalyavin}}, \bibinfo {author}
  {\bibfnamefont {S.}~\bibnamefont {Langridge}}, \bibinfo {author}
  {\bibfnamefont {V.}~\bibnamefont {Saidl}}, \bibinfo {author} {\bibfnamefont
  {P.}~\bibnamefont {Nemec}}, \bibinfo {author} {\bibfnamefont {A.~W.}\
  \bibnamefont {Rushforth}}, \bibinfo {author} {\bibfnamefont {B.~L.}\
  \bibnamefont {Gallagher}}, \bibinfo {author} {\bibfnamefont {S.~S.}\
  \bibnamefont {Dhesi}}, \bibinfo {author} {\bibfnamefont {F.}~\bibnamefont
  {Maccherozzi}}, \bibinfo {author} {\bibfnamefont {J.}~\bibnamefont
  {{\v{Z}}elezn{\'y}}}, \ and\ \bibinfo {author} {\bibfnamefont
  {T.}~\bibnamefont {Jungwirth}},\ }\bibfield  {title} {\enquote {\bibinfo
  {title} {Antiferromagnetic structure in tetragonal cumnas thin films},}\
  }\href {\doibase 10.1038/srep17079} {\bibfield  {journal} {\bibinfo
  {journal} {Scientific Reports}\ }\textbf {\bibinfo {volume} {5}},\ \bibinfo
  {pages} {17079} (\bibinfo {year} {2015})}\BibitemShut {NoStop}%
\end{thebibliography}%

\end{document}